\documentclass[useAMS,usenatbib]{mnras}
\usepackage{epsfig}
\usepackage{epstopdf}
\pdfoutput=1 

\usepackage{times,txfonts}
\usepackage[T1]{fontenc}
\usepackage{ae,aecompl}

\usepackage{graphicx}	
\usepackage{amsmath}	
\usepackage{amssymb}	

\usepackage[usenames,dvipsnames]{color}

\newcommand\ocen{$\omega$\,Cen}

\newcommand\Msun{{M$_\odot$}}
\newcommand\Thbb{{T$_{\rm HBB}$}}

\newcommand\msun{\rm{M$_{\odot}$}}

\newcommand\mmass{M$_{\rm mass}$}

\newcommand\teff{T$_{\rm eff}$}
\newcommand\thbb{T$_{\rm HBB}$}
\newcommand\Tc{T$_{\rm c}$}

\title[Multiple populations in globular clusters]{ A unique model for the variety of  multiple populations formation(s) in globular clusters: a temporal sequence}

\author[F. D'Antona et al.]{
F. D'Antona$^{1}$, E. Vesperini$^{2}$,  A. D'Ercole$^{3}$,  P. Ventura$^{1}$, A.P. Milone$^{4}$,  
\newauthor A. F. Marino$^{4}$   \& M. Tailo$^{1}$ \thanks{E-mail: franca.dantona@gmail.com (FD)
}  
\\
$^{1}$INAF, Osservatorio Astronomico di Roma, Via Frascati 33, 
I-00078 Monteporzio Catone (Roma), Italy.\\
$^{2}$ Department of Astronomy, Indiana University, Swain West, 727 E. 3rd Street, IN 47405 Bloomington (USA)\\
$^{3}$ INAF, Osservatorio Astronomico di Bologna, via Ranzani 1, I-40127 Bologna, Italy\\
$^{4}$  Research School of Astronomy \& Astrophysics, Australian National University, Canberra ACT 2611, Australia\\
}
\date{Accepted XXX. Received YYY; in original form ZZZ}

\pubyear{2015}

\begin{document}
\label{firstpage}
\pagerange{\pageref{firstpage}--\pageref{lastpage}}
\maketitle

\begin{abstract}
We explain the multiple populations recently found in the `prototype' Globular Cluster (GC) NGC\,2808  in the framework of the asymptotic giant branch (AGB) scenario. The chemistry of the five --or more-- populations is approximately consistent with a sequence of star formation events, starting after the supernovae type II epoch, lasting approximately until the time when the third dredge up affects the AGB evolution (age $\sim$ 90--120Myr), and ending when the type Ia  supernovae begin exploding in the cluster, eventually clearing it from the gas. 
The formation of the different populations requires episodes of star formation in AGB gas diluted with different amounts of pristine gas. 
In the nitrogen--rich, helium--normal population identified  in NGC~2808 by the UV Legacy Survey of GCs, the nitrogen increase is due to the third dredge up in the smallest mass AGB ejecta involved in the star formation of this population. The possibly-iron-rich small population in NGC~2808 may be a result of contamination by a single type Ia supernova.\\
The NGC\,2808 case is used to build a general framework to understand the  variety of  `second generation' stars observed in GCs. Cluster-to-cluster variations are ascribed to differences  in the effects of the many processes and gas sources which may be involved in the formation of the second generation. We discuss an evolutionary scheme, based on pollution by delayed type II supernovae, which accounts for the  properties of  s-Fe-anomalous clusters. 
\end{abstract}

\begin{keywords}
globular clusters: general -- globular clusters: individual: NGC 2808,  NGC 1851, NGC 5286, NGC 6121, NGC 6656 -- stars: formation -- supernovae: general --binaries: close
\end{keywords}


\section{Introduction}
\label{Intro}
The Hubble Space Telescope UV Legacy Survey of Galactic GCs  \citep{piotto2015}, is exploiting the sensitivity of UV photometric observations to different molecular bands to disentangle GCs' multiple stellar populations  in 57 clusters.   
Among the surprises of the UV Survey's first results, there is the discovery that the multiple-population cluster NGC~2808 not only contains the three populations known to differ in helium content \citep{dantona2005,piotto2007} and in light element abundances distribution \citep{carretta2006}, but it includes at least two more families \citep{milone2015}. The five populations identified have been labeled by \cite{milone2015} with letters from A to E. The intermediate-- and the high--helium main sequences (MS) are easily identified and correspond to populations D and E, respectively. The population with almost standard helium includes at least three stellar groups, namely A, B, and C. In particular,  group C includes a significant fraction of cluster stars  ($\sim$25\%), does not differ in helium content from group B, but shows the fingerprints of a high surface nitrogen content. Group A is a small population ($\sim$6\%) for which spectroscopic data are not available and which could be slightly more metal rich than the rest of the cluster stars. Consequently, group B alone is fully compatible with a ``first generation" (FG).\\
Very recently \cite{carretta2015} has reanalyzed the large database of spectroscopic data collected in several years for the cluster giants, and has shown that these five populations can be distinguished also from the spectra, thanks to their clustering in the Mg--Na, O--Mg Mg--Si, and O--Na planes.
The groups identified spectroscopically and photometrically do not completely overlap: spectroscopy can distinguish two different groups among the stars of each of the photometric groups B, C and D (which  include the spectroscopic groups P1, P2,  I1 and I2), while the E group is the same (also in nomenclature) in the photometric and spectroscopic definition, but also notice that  the bayesan analysis of  the photometric groups \citep{milone2015} has shown that both populations B and C may host distinct subpopulations. There are no spectroscopic data for the stars of group  `A'. 
In summary, the comparison between spectroscopic and photometric data suggests that there might be seven  separate groups in this cluster.

NGC\,2808 is sometimes considered a `prototype' cluster for multiple populations, as it has been one of the first clusters, together with \ocen, in which helium variations have been postulated and discovered. 
NGC\, 2808, however, in addition to populations with milder chemical anomalies similar to those found in most GCs, hosts also a population with more extreme chemical variations, found in only few other clusters. In this work, our goal is to explore the viability and the key ingredients of a unified model for the origin of globular cluster abundance patterns, capable of explaining both the more modest and widespread anomalies and the more extreme ones.

Can the models, built to account for a simpler scheme of multiple populations, be extended to deal with the more complex observational findings emerging from recent studies? 
In the latest 10-15 years, the observations promoted the formulation of several different theoretical explanations and provided  a large number of  constraints that have challenged the proposed theories.

We emphasize here that, currently, the strongest observational constraints come from the chemical abundance properties and patterns revealed by spectroscopic and photometric observations. Any effort aimed at  identifying  the fundamental ingredients in the formation of multiple populations should be driven by these observational constraints which provide fundamental clues on the possible sources of gas out of which SG stars formed. While many factors related to a cluster internal dynamics and its complex interplay with the external galactic environment (particularly in the early stages of galaxy formation) have certainly played a key role in determining the current properties of multiple-population clusters, the first step in assessing the viability of different models should be guided by the observed chemical properties. Although it is extremely important to explore all the implications of the presence of multiple stellar populations for a cluster dynamical history, using other constraints, related to much more uncertain aspects of the formation history of globular clusters and of their host galaxies, to rule out any model appears premature.

We will exploit the AGB scenario \citep{ventura2001, dercole2008, bekki2011} to interpret the five populations of NGC\,2808. This scenario has been developed with sufficient detail to address the main photometric and spectroscopic constraints. Moreover we will provide a critical analysis of the other main scenarios for the formation of the distinct stellar populations at the light of the recent observations on NGC\,2808.

The AGB scenario could nicely and simply explain helium differences among the stars in this cluster, which had been predicted by analyzing  the horizontal branch morphology \citep{dc2004}. When the triple MS was discovered \citep{dantona2005, piotto2007}, parametric modeling of the helium evolution in the cluster became more detailed, as  first shown in the hydrodynamical computations in \cite{dercole2008}, and subsequently in simple chemical evolution models \citep{dercole2010, dercole2012} including other chemical abundance patterns.  This same model, slightly extending the time of formation of the multiple population, is here shown to provide a good explanation for the other two populations. The 5 populations, in the order BEDCA, are shown here to represent the outcome of a clear temporal sequence of star formation events. 
Specifically, we show that the different chemical fingerprints are in fact in close relation with the temperatures at the bottom of the convective envelope, \Thbb, of the AGB stars providing  the gas out of which SG stars formed. The decrease of \Thbb\  with the evolving initial mass reveals  the possible temporal sequence for the formation of the different observed patterns. 

In the specific case of NGC~2808, populations C and A simply result from a slight extension in the duration of the star formation epoch proposed in the context of  the AGB model for populations B, E and D by \cite{dercole2008, dercole2012}.

The AGB scenario is able to trace the star formation history (SFH) of multiple populations in NGC 2808. This model can be easily adapted to any other GC, once we allow for simple--minded differences in the events which modulate SFH in each cluster, based on the same ---or similar,  depending on the metallicity--- temporal evolution of the polluting material. In this respect, this work aims at providing a reasonable baseline to understand this complex problem.

In this paper we explore the possible chemical evolution of NGC~2808 with a simplified model, discuss the possible events separating the star formation epochs of the five (or more) populations, and suggest which parameters (e.g. the duration of the SG star formation phase, the timing and intensity of the pristine gas accretion phase, the properties and frequency of the dividing events, the role of delayed binary type II supernovae, the sudden or slow onset of SN~Ia explosions) may produce different abundance patterns in other clusters. 

The outline of the paper is the following.\\
In \S~\ref{observations} we summarize the HST spectrophotometric and the spectroscopic data which form the basis of our knowledge about the 5 (or more) populations hosted in NGC~2808. We show that the magnesium variations are at variance with the results of most models proposed so far, while they can be compatible with the AGB scenario. We anyway describe the difficulties from AGB nucleosynthesis, and the possible solutions. \\
In \S~\ref{sec:3} we describe how the yields of AGB ejecta vary with time, due to the decrease of the HBB temperatures with decreasing initial mass, and discuss that the patterns of abundances may be correctly described only if, at some stage(s) the ejecta are diluted with pristine matter.
In \S~\ref{sec:4} we produce a parametric model for NGC~2808, able to explain ---at least qualitatively--- the evolution with time of the composition of the multiple populations. The time order of populations A, B, C, D and E in \cite{milone2015} is given by the acronym BEDCA. In this timeline, we propose feasible explanations for the new populations C and A. In \S~\ref{HB} we show that group C stars are also compatible with their location in the horizontal branch. \\
In  \S~\ref{sec:metals} we present a brief discussion concerning the chemical patterns of clusters of different metallicity and how they are consistent with the trends in the AGB yields as a function of the model metal content. \\
In \S~\ref{frame} we extend the model to other clusters, after examining in detail the different epochs during the early life of clusters, in terms of both AGB ejecta composition and of events perturbing the star formation. The possible role of delayed type~II supernovae (SN~II) and type~Ia supernovae (SN~Ia) in affecting the resulting chemical properties of second-generation stars is discussed.
Finally in \S~\ref{sec:delSNII}  we propose a model to explain the features of s-Fe-anomalous clusters, and we elaborate on the possible iron pollution of both kind of late supernovae. Results are summarized in the table \ref{tab:table3} in the Conclusions section.

\section{Models versus observational constraints}
\label{observations}

\begin{table}
	\centering
	\caption{Spectroscopic and spectrophotometric abundances for NGC 2808 populations}
\center
\scriptsize {
\begin{tabular}{ccccccc}
\hline
\hline
 C15$^1$  & M15$^2$ & [O/Fe]$^3$ &  [Mg/Fe]$^3$  &  [Al/Fe]$^3$  & [Na/Fe]$^3$  & [Fe/H]$^3$  \\\\
  P1       &   B &        0.30  &     0.38 &  0.05  &    0.03  & $-$1.13  \\
\hline  
  && $\Delta$[O/Fe]\,$^4$  & $\Delta$[Mg/Fe]&  $\Delta$[Al/Fe] & $\Delta$[Na/Fe]  & $\Delta$[Fe/H] \\ 
P2,I1 &   C &    $-$0.1  &        0.0  &  0.2    &   0.2   &  0.0   \\
I2       &   D &    $-$0.7  &    $-$0.3\,$^5$  &  1.0\,$^5$   &   0.4   &  0.0  \\
E       &  E &    $-$0.9  &   $-$0.4\,$^5$  &  1.2\,$^5$   &   0.8   &  0.0  \\
\hline
\hline
\end{tabular}
}
\center
\scriptsize {
\begin{tabular}{lccccc}
\hline
\hline
Pop.   &  $\Delta$Y &  $\Delta$[C/Fe] &  $\Delta$[N/Fe] & $\Delta$[O/Fe] \\
\hline
C      &    0.00    & $-$0.3 &    0.6  & $-$0.1    \\
D      &    0.05    & $-$0.7 &    0.9  & $-$0.5    \\
E      &    0.10    & $-$1.0 &    1.2  & $-$0.8    \\
\hline
\hline
\end{tabular}
\label{tablemilone}     
}
\\
$^1$\, C15: Carretta 2015 groups definition  \\
$^2$\, M15: Milone et al. 2015 groups definition \\
$^3$\, values from \cite{milone2015}, matched to \cite{carretta2006} and \cite{carretta2014}\\
$^4$\, all variations with respect to B group values, are from \cite{milone2015} apart from those in ($^5$)\\
$^5$\, from \cite{carretta2015} 
\end{table}

\subsection{Observations}
\label{sec:obs}
\cite{carretta2015} summarizes the patterns of abundances of light elements which define the spectroscopic  evidence for the presence of multiple populations in NGC~2808. Variations are found in the `classical' elements oxygen, sodium, magnesium and aluminum, but also in the heavier element silicon. Further, variations in potassium \citep{mucciarelli2015} must be also considered.  In this work, Carretta also compares his spectroscopic analysis with the photometric results, and remarks in details the specificity of the two different approaches. 

One important observational clue is that magnesium decreases by about 0.4~dex, and silicon increases by about 0.2~dex  between the  P1 (primordial 1) population and the `extreme' E population \citep{carretta2015}.

The photometric two color diagram introduced by \cite{milone2012} and used to study the populations of NGC 2808 in \cite{milone2015} works on the amplification of the differences in C, N and O abundances in the spectral distribution of stars, which is obtained when using specific combination of the UV and optical HST passbands, e.g. the plane of the color  y=m$_{F336W}$-m$_{F438W}$\ versus the color  x=m$_{F275W}$-m$_{F814W}$.
As C, N and O are the main drivers of the clustering of stars in this plane, we will refer to this diagram as `the CNO--two--color diagram'.  
Here, the stars group into at least 5 main regions, labelled from A to E. 

The groups B, D and E are easily identified as the helium normal, intermediate and extreme groups already known. Two more groups are present: C, nitrogen richer than B, and the small group A.
Spectrophotometric differential analysis of the magnitudes in the different bands from \cite{milone2015} provides the results in Table\,\ref{tablemilone} for the populations C, D and E with respect to group B assumed to be the first generation.  
The comparison  with the \cite{carretta2015} analysis allows to include in the table the differential values for Mg and Al. 
Errors are not listed in the table, as the data come from non homogeneous measurements, for which a rigorous evaluation is difficult. We use the abundance differences mostly to show the qualitative  agreement of their values with the model trends (Figs. 3 and 4).

The A group is not listed in the table, as the differential analysis in \cite{milone2015} has been done by assuming an helium abundances smaller than in the reference group B by $\delta$Y=--0.03, while in this work we assume that A group stars are more iron rich by $\delta$[Fe/H]$\sim +0.1$,  a value estimated by the slightly cooler location of the A with respect to the B MS \citep{milone2015}. Notice that the small color increase may also be due to an increase in the specific abundances of oxygen, or silicon, or sulfur, as shown in the detailed study by \cite{vandenberg2012}. A conclusive understanding will require abundance measurements in the atmosphere of these stars.
 
 \subsection{Constraining viable models with Mg and Si}
 \label{sec:prequalif}
 In NGC~2808 there is evidence of both Mg depletion  and Si enhancement. The maximum Mg depletion, about 0.4~dex,  is found among the stars of population E in \cite{carretta2015}. This fact alone is sufficient to rule out all proposed models for the formation of multiple populations based on the nucleosynthesis product available from H-core burning in massive stars, namely, the Fast Rotating Massive Star model \citep[e.g.][]{decressin2007}, the Massive Interacting Binary model \citep{demink2009}, or the accretion model \citep{bastian2013} for which the nucleosynthesis products of  massive binaries are the source of chemical anomalies. \\
Inside these stars, the maximum temperature does not exceed $\sim$65MK, an extreme value reached only during the latest phases of core H--burning by the most massive models  \citep{decressin2007},  so magnesium can not be affected by proton captures and silicon can not be synthesized. The problem is discussed in the \cite{prantzos2007} analysis, based on one--zone chemical evolution models at fixed temperature, which are not linked to any specific computation. 
Notice that this constraint can not be changed by variations in the details of the stellar models, as it depends on the fundamental properties of stellar structure. By combining the observed mass--radius relation, for stars burning core hydrogen via the CNO cycle, with the hydrostatic equilibrium plus the perfect gas law, the central temperature \Tc\ versus total mass relation is described by a power law with a very shallow exponent $\sim$0.2--0.3, providing values of $\sim$50MK for the interior temperature of stars at M$\sim$100\Msun. \\
In an attempt to increase the temperature, in the H--core burning phase, up to the canonical 75MK which allow Mg and other advanced p--capture reactions in the interiors, \cite{denissenkov2014} have suggested that supermassive stars are responsible for the formation of SG stars in GCs.  However such stars have no observational counterparts.

In conclusion the only possibility for massive star models to overcome this fundamental problem is to assume a larger cross section for the $^{24}$Mg(p,$\gamma$)$^{25}$Al reaction, possibly through a not yet identified resonance at the needed temperatures. An increase in the cross section by a factor 1000 is, however needed  \citep{decressin2007}, and because of such a large discrepancy we think these models can be ruled out based on what is now known from the nuclear evolution point of view.
\cite{renzini2015} further discuss the inability of all these models to produce  the observed discreteness and to explain cluster-to-cluster differences in the properties of multiple populations, as well as the reliability of supermassive star models as polluters for the multiple populations, but here we rule them out on the simple basis that they can not be used to describe the abundances observed by \cite{carretta2015}, {\it et de hoc satis}.

\subsection{The problems of nucleosynthesis in the AGB scenario}
\label{sec:agbnuc}
So we claim that  p--captures on  Mg nuclei and other advanced nucleosynthesis products find a place for the nucleosynthesis in the hot bottom burning (HBB)  envelopes of massive AGBs, in the models in which the temperature of HBB (\thbb) is sufficiently high  \citep[$\gtrsim10^8$K,][]{prantzos2007, ventura2011}. Nevertheless, the {\it quantitative} result of Mg burning in massive AGBs of metallicity adequate to describe NGC~2808 also faces problems: in the models that will be used here \citep{ventura2013} the maximum Mg depletion is $\sim$0.17 dex, while the data require $\sim$0.4 dex. 
We have discussed in specific models \citep{ventura2011} that the \thbb's of the massive AGB are indeed large enough to allow the reaction  $^{24}$Mg(p,$\gamma$)$^{25}$Al to occur, but {\it longer} evolutionary times in AGB are needed to get a quantitative agreement in the total depletion. This can be achieved by assuming a {\it smaller} mass loss rate. 
The  essence of  the problems with AGB modeling is the following.
{\it Smaller mass loss rates than those adopted in these models allow magnesium and aluminum to be processed more efficiently,  but have the counter effect of reducing the total yield of sodium, disfavoring the O--Na anti correlation}. This trend is also confirmed in the models by \cite{doherty2014}, which show a larger Mg depletion, but also full depletion of sodium in the yields. 
In fact, the nucleosynthesis in HBB  has to face the plain fact that sodium is destroyed at the same time oxygen is destroyed, while the abundances displayed by second generation stars in GCs require high sodium abundances at low oxygen abundances. 

In the following we list the three quantitative discrepancies between AGB nucleosynthesis yields and observed abundance patterns that are solved if we allow for smaller mass loss rates. 

{\bf 1.} The oxygen depletion in the most massive super--AGBs is not large enough. To address this problem, we have invoked deep mixing in the giant progenitors of the most extreme population, a mixing which should be favored by the large helium abundance of these stars \citep{dantonaventura2007}, and we adopt this solution also here (see \S~\ref{oxygen}). Nevertheless, smaller mass loss rates in the super--AGB phase would lead to the required large depletion of oxygen and would provide an additional possible solution to this problem. 

{\bf 2.} Apart from $^{24}$Mg depletion, some other HBB proton captures require a longer time than allowed by the mass loss rates chosen in our \cite{ventura2013} paper and in the previous computation. In particular, the silicon increase by proton captures on aluminum  is found in \cite{ventura2013} at the level of $\sim$0.05 dex, while $\sim$0.15 dex are required by \cite{carretta2015} data for NGC~2808. Also in this case, a larger production requires longer evolutionary times. 

{\bf 3.}  Taking at face value the modest $\delta$Y implied by the observational width of the MS of some clusters in recent observations,
\cite{bastian2015}  found a discrepancy between the range covered by data in O--Na plane and the range allowed by models.
 Below, we discuss some limitation of this analysis, and other possible ways to overcome this problem. However, taking this result at face value, also in this case the difficulty is removed {\it if the models have a larger oxygen depletion for a given \thbb}, and this can be achieved assuming smaller mass loss rates during the AGB evolution. If models with smaller \thbb\ can describe well the O--Na data, they are progeny of lower initial masses and have smaller Y yield, as required by the comparisons in \cite{bastian2015}, solving the discrepancy. 

The mass loss rates in the AGB models by \cite{ventura2013} were originally calibrated on a comparison with the luminous Lithium rich stars in the Magellanic Clouds \citep{ventura2000}, but this calibration is also able to provide a {\it positive} sodium yield, in spite of the contemporary burning of sodium and oxygen. If we allow for a mass loss rate smaller, say  by a factor three ---that is, we triple the time for p--captures in the AGB envelope--- we may find a solution for the four problems listed above, {\it at the expenses of burning too much sodium} \citep[see, e.g.][]{ventura2011, doherty2014}.  \\
The problem of the sodium destruction is examined in detail in \cite{renzini2015}, where it is suggested that a reduction by a factor five of the reaction rate  $^{23}$Na(p,$\alpha$)$^{20}$Ne  would allow sodium to be destroyed at a slower rate, so that its global yield may be in the range of the observed sodium abundances also in the stars in which oxygen is largely depleted. So the discrepancies with observations  of AGB nucleosynthesis can be solved by adopting a reduced mass loss rate, if the  proton capture rate on sodium is smaller.

Reduction of the sodium burning rate also mitigates the problem of the sodium yield in very low metallicity models. 
Very low metallicity clusters, as  \citep[e.g. M~15,][]{carretta2009a},  show the typical anticorrelation O--Na. As opacities are smaller at lower metallicity, the \thbb's of the AGB models are larger, so that sodium is burned more efficiently, and it is even more difficult than for clusters of larger metallicity to preserve a large sodium yield \citep{ventura2009}. This problem might show up particularly in NGC~2419 \citep{ventura2012ngc2419}, because this cluster contains a large, extreme, population with very high helium, interpreted as a population born from pure ejecta \citep{dicrisci2011,dicrisci2015}. In this cluster, anyway, there are very few measurements of sodium abundances \citep{cohenkirby2012} and the presence of a O--Na anti correlation is not clearcut \citep{ventura2012ngc2419}. 

The  truly needed range of reduction in the rate of  $^{23}$Na(p,$\alpha$)$^{20}$Ne, beyond the limits published by \cite{hale2004}\footnote{The lower limit of the cross section determined by \cite{hale2004} is only 20--25\% smaller than the recommended rate, at T$\sim$100\,MK.}, should  be verified by model computation, together with a check of the best mass loss rate producing agreement with the whole set of abundances displayed by SG stars. We estimate that the reduction required should indeed be in the range 2--5. This reaction proceeds populating several states. Low-energy resonances in the $^{23}$Na+p system have been studied by \cite{zyskind1981}, \cite{goerres1989} and \cite{rowland2002}, down to the states of energy of about 200 keV. Below the lowest observed resonance at E$_{\rm cm}$ =170\,keV, there are four other states, which could correspond to resonances at 5, 37, 138, and 167\,keV.  \cite{hale2004} describe an indirect experiment to determine spectroscopic factors via ($^3$He, d) for these low energy states, and concluded that the state at E$_{\rm cm}$=138\,keV may make a significant contribution to the rates of $^{23}$Na(p,$\alpha$)$^{20}$Ne . Due to the weak experimental constraints, the uncertainty of the reaction rate at the relevant astrophysical energy (70-110 MK) is very large and could be larger than a factor 2.
To shed light to the intensity of this state, several experiments are planned in different laboratories, e.g. LUNA \citep{luna2009} which is concentrating on the gamma channel  $^{23}$Na(p,$\gamma$)$^{24}$Mg. 

In the present work we will mostly consider the {\it trends with time} of the yields predicted in AGB models, as we are interested in qualitative or semi-quantitative comparisons. The important point of the comparison is to understand whether these trends, which depend primarily on \thbb, provide a general scheme to understand the different populations in NGC~2808, and if this scheme can be simply adapted to other clusters.   
 
\subsection{Helium versus p-capture elements}

As discussed in the previous section, a reduction of the proton capture rate on sodium, together with the assumption of smaller mass loss rates, can solve all the quantitative discrepancies between the AGB yields and  the observed abundance anomalies; in particular, a larger depletion of oxygen for each evolving HBB model alleviates or cancels the helium discrepancy presented by \cite{bastian2015}. Although we think this is a likely solution to all the discrepancies, we further discuss here the helium problem raised by \cite{bastian2015}.

It is important to emphasize that the all the $\delta$Y are ``indirect" determinations from the observations. In particular,  the values quoted in \cite{bastian2015} depend on a number of assumptions on how the colors of main sequence stars must be interpreted.
The helium lines are visible only at \teff$\gtrsim$8000K, and the abundance measured in the spectra (and also here we have to rely on models) is directly linked to the original abundance in the star only for a very small range of \teff\ in the horizontal branch \citep{behr2003, monibidin2007}. 

Besides, a careful look at the data adopted by \cite{bastian2015} is necessary to assess the real importance of the problem. In particular, the O--Na data in \cite{bastian2015} are mostly taken from the useful collection by \cite{roediger2014},  joining data from different sources.  As an example, the data of the cluster 47~Tuc used in the work are the combination of two data sets having different zero points of abundances. \\
A most critical point is that the percentage of stars along the predicted dilution curves should be considered in detail. In fact, the $\delta$Y in \cite{milone47tuc2012} is based on the {\it average} distance in MS color between the two populations, and is certainly not the maximum allowed value.
A longer discussion of this issue is out of the scope of the present work, but it is under detailed scrutiny using the UV Legacy Survey data (Milone et al., in preparation).

Another factor which may affect this possible discrepancy is the following: the helium yield from massive AGBs is the result of the second dredge up (2DU), occurring between the H--rich envelope of the giant and the H--exhausted core before the model climbs up the AGB (see \S~\ref{thbb-time}). On the contrary, the p--processing occurs at a later stage,  during the HBB in the AGB envelopes, and is highly dependent on entirely different physical inputs (we have discussed mass loss, but one of the most critical is the convection modeling, see \citet{renzini2015} for a longer discussion and references). Thus, should  further observations confirm this discrepancy, the solution to this problem might be in models with different modalities of the 2DU, reducing the discrepancy in the $\delta$Y. In Table~\ref{tab:table1} we show how a simple difference in the models alters the envelope Y abundance after the 2DU by $\delta$Y=0.02 for masses 6 and 5\Msun. Notice that the C-O core mass is also altered, so a fully consistent modeling is required for a global study of this problem. The second column in Table~\ref{tab:table1} lists the value of the parameter $\zeta$\,  describing the overshooting in our models \citep{ventura1998}. $\zeta$ is the e-folding distance ---in units of the thickness of the convective region expressed as a fraction of the pressure scale height H$_{\rm p}$--- over which the model assumes that the convective velocities decay exponentially from the formal convective border. A parameter $\zeta$=0.02 is the value which is generally adopted in the models, based on the comparison with observations of the main sequence width, but larger and or smaller values can not be excluded.

\begin{table}
	\centering
	\caption{Helium mass fraction at the 2DU as function of the extra mixing during core--H burning}
	\begin{tabular}{cccc}     
		\hline
		M$_{\rm AGB}$/\msun$^1$  & $\zeta^2$ &  M$_{\rm core}$/\msun$^3$  &   Y$_{\rm 2DU}^4$ \\
		\hline
6.0  & 0.01  & 0.988  &  0.344 \\
6.0  & 0.02  & 1.025  &  0.349 \\
6.0  & 0.03  & 1.060  &  0.362 \\
\hline
5.0  &  0.01 &   0.904 &   0.324 \\
5.0  &  0.02  &  0.917  &  0.332 \\
5.0  &  0.03  & 0.939   &  0.343 \\
		\hline		
		\end{tabular}
\\	
$^1$ Initial mass - Y=0.25, Z=0.001 and [$\alpha$/Fe]=0.4\\
$^2$ e-folding distance of overshooting (Ventura et al. 1998)
 \\
$^3$  C-O core mass at the AGB\\
$^4$  Helium mass fraction after the 2DU
	\label{tab:table1}
	\end{table}
In conclusion, we think that, at this point, the claim by  \cite{bastian2015} that the constraints  from O--Na--Y data rule out the  AGB model is hasty; until  additional theoretical and observational work on this issue is carried out and further light on the actual extent of the problem and its possible solutions is shed, much caution in drawing any strong conclusion should be exercised.

\section{Ingredients of the AGB model}
\label{sec:3}

\subsection{The AGB ejecta composition versus time}
\label{thbb-time}
In the previous sections we discussed that several modifications to the current AGB models will be needed, and that future observations will be critical in driving the refinement of the AGB models \citep[see also][]{renzini2015}. Now we focus our attention on whether the fundamental ingredients of the AGB scenario may satisfy the main chemical constraints displayed by multiple populations. 

\begin{figure}
\centering{
\includegraphics[width=8.5cm]{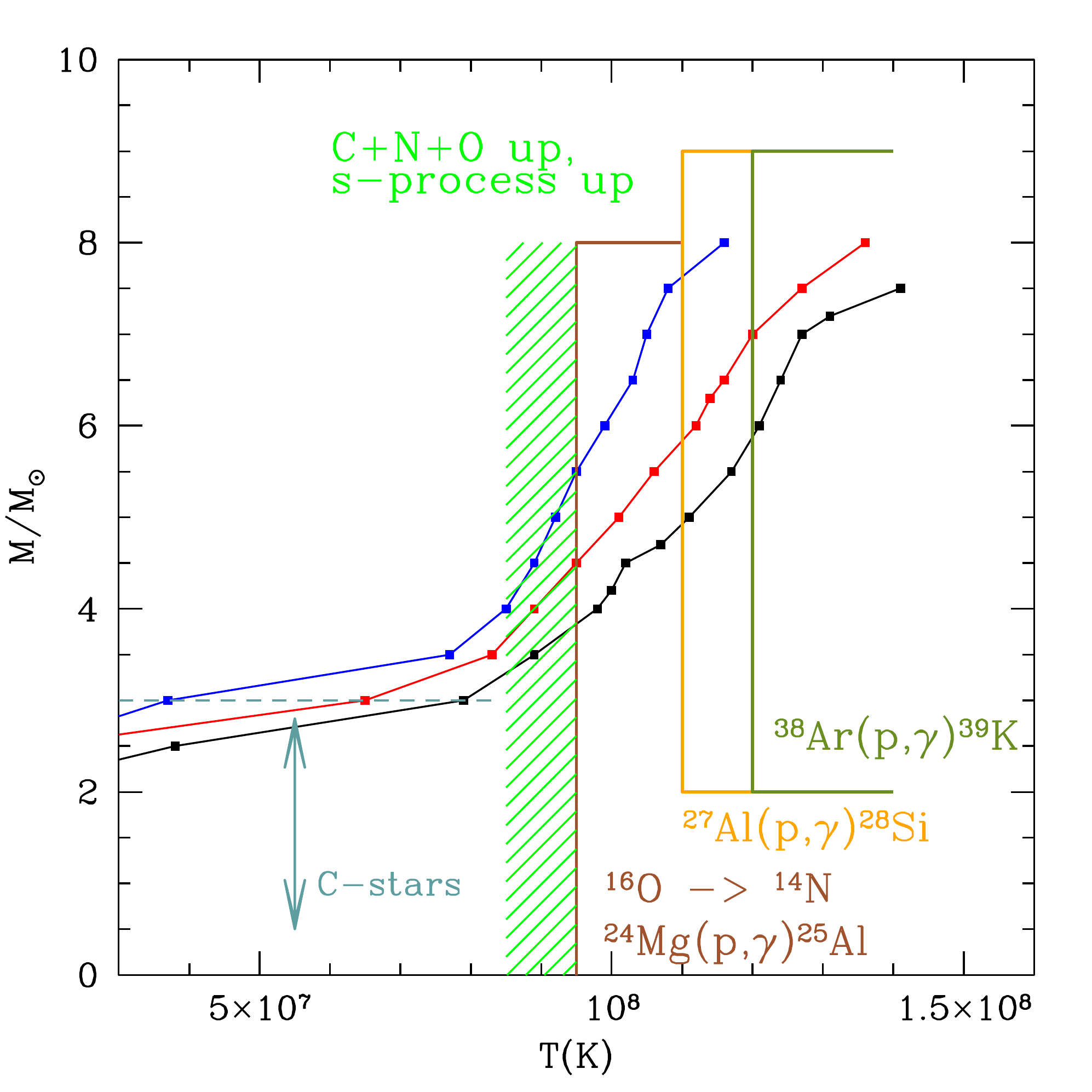}
\caption{Initial mass versus  the (maximum) temperature at the bottom of the convective envelope for masses evolving in AGB, from Table 3 in  Ventura et al. (2013), for metallicities Z=0.008 (leftmost, blue) 0.001 (center, red) and 0.0002 (right, black). The temperatures above which some important nuclear reactions (labelled) are efficient are shown as vertical lines. The dashed green region indicates the temperatures at which the 3DU up becomes so relevant that envelope ejecta have C+N+O and s--process larger than the values of the starting composition. The dashed (horizontal) line indicates the approximate mass at which the 3DU leads to form Carbon stars, as the conversion of C into N becomes inefficient.}
\label{ff1}
}
\end{figure}

During the thermally pulsing phase, models of large enough mass reach the very high temperatures, \Thbb, necessary to achieve the proton captures which modify the light elements abundances previously quoted. Fig.\,\ref{ff1} shows  the maximum  \Thbb\ reached, as a function of the initial mass and metallicity, in the evolutionary computations  by \cite{ventura2013}. The assumptions behind these results can be found in their paper. Here we simply remind the reader that the relation of \Thbb\ versus the initial mass depends on two main parameters: 

1) the assumptions made on core overshooting during the H-core burning, which provides the mass of the H--exhausted core. This He--core mass is modified by the 2DU \citep{beckeriben1979}, and has a smaller reduction for stronger core--overshooting assumptions (see Table~\ref{tab:table1}).  The mass of the C--O core remnant at He--exhaustion \citep{bressan1993} is very close to the He--core mass after the 2DU. Larger C-O mass means larger \thbb \citep[e.g.][]{bs1993}.

2) the assumptions made for the efficiency of external convection determine \thbb\ for each core mass \citep{ventura2005a}.

Different choices for these parameters change the temperature values in Fig.~1. In particular, a smaller efficiency of convection may not allow to reach \thbb's above 10$^8$K, or it may be reached only at very low metallicity, or only for the largest masses. These models would not be able to explain, e.g., the formation of silicon. In addition, C+N+O would increase in the whole mass range of massive AGBs, providing results at variance with observations (Fenner et al. 2004).

While we must take  Fig.~1 with caution for what concerns the temperature numerical values, the trends with metallicity and mass have a validity beyond the particular assumptions of specific models. Notice that the \cite{ventura2013} models follow a line of models computed with very efficient convection \citep{cm1991, cgm1996}, so they probably already have the largest \thbb\ ---at fixed core mass--- that can be reached in 1D models. On the other hand, these models use a moderate overshooting, so a bit larger core mass (and a larger \thbb) is possible for a given initial mass in models with increased overshooting. 

The role of metallicity in determining the possible HBB processing is very clear: the smaller the metallicity, the larger is \thbb\ (due to the smaller surface opacities). This means that the products of the hottest HBB (e.g. Si production from proton captures on Al) can be found only at low metallicity, and for large initial masses (see also \S~\ref{sec:metals}).

The final nucleosynthesis products depend on \thbb, but also on the total time spent in the AGB phase, and the total time spent in the AGB phase, in turn, ultimately depends on the mass loss rate. High mass loss rate may drastically limit the p--captures \citep{ventura2005b}. In fact, part of the differences between the sodium and oxygen yields provided by \cite{ventura2013} and \cite{doherty2014} in the massive AGB star regime are due to the different prescriptions adopted for mass loss.  \\
This problem is even more evident in the super-AGB computations by \cite{ventura2011sagb}, where the algorithm implemented for the mass loss rate \citep{blocker1995} leads to {\sl smaller} oxygen depletion than in the most massive AGB models\footnote{This does not happen in the SAGB models by \cite{siess2010} and \cite{doherty2014}, which show a very large oxygen depletion, but these same models have the (usual) problem that the sodium is destroyed instead of being enhanced ---as discussed in section 2.3}. Unfortunately, there are no observational tests for constraining the mass loss in super-AGBs.

Fig.~1 shows the boundaries in \Thbb\ necessary to process some important elements by proton capture. The boundary for the operation of the full CNO cycle, with an efficient conversion of oxygen into nitrogen, is at about 90--95\,MK. At $\sim$100\,MK the proton captures on $^{24}$Mg begin to be efficient\footnote{This temperature looks much larger than the T$\sim$75\,MK quoted above as temperature for Mg burning. Notice that we {\it are not} dealing here with stellar interiors: the densities at the bottom of the convective envelope ($\sim$10--15 g cm$^{-3}$) are comparable or larger than the density in massive stars interiors, but the timescale for the entire AGB evolution is shorter.}. The production of silicon requires even larger \Thbb$\sim$110\,MK, and the possible potassium production advocated by \cite{ventura2012ngc2419} requires \Thbb$>$120\,MK. In addition, adjustments of the relevant cross sections for these proton captures is required in the models. 

In Fig.~1, below 10$^8$K, we highlight as a green shaded area the important regime in which oxygen is still affected by p-captures, and the lifetimes in AGB increase. These models begin to show a strong effect of the third dredge up (3DU), the process by which the external convective region reaches stellar layers previously affected by efficient 3$\alpha$\ burning after a thermal pulse. Thus carbon from the He--inter-shell is convected to the surface, and then converted into nitrogen by HBB. So {\it in this range the total C+N+O content, and the s-process elements abundance of the stellar ejecta increase.} 
Also the amount of CNO enrichment depends on uncertain stellar parameters; in particular it strongly depends on the duration of the AGB phase, and thus on the mass loss rate. Smaller mass loss rate ---either due to the formulation adopted, or to the choice of free parameters, or due to the different stellar parameters from which the mass loss function depends (e.g., the stellar radius or \teff, which in the giants is linked to the atmosphere computation)--- means longer phase duration, a larger number of thermal pulses and 3DU events, and a larger CNO enhancement. Finally, also the numerical method adopted to compute the 3DU has an influence on the final outcome\footnote{For intermediate mass AGBs, the 3DU is actually a spontaneous event \citep{ir1983}, which occurs in the stellar structure without any further assumptions. In smaller mass stars, it is generally achieved through a further parametrization of envelope boundary overshooting.}. Thus, even more than for the other yields, our results for the total CNO enhancement are meant to provide approximate indications rather than strict predictions. \\
Nevertheless, one firm result is the following: if star formation using up AGB ejecta goes on for a {\it long} time ---where this precise time depends on the models adopted, and in our models is $\sim$100\,Myr from the formation of the FG, or $\sim$60\,Myr from the end of the SN~II epoch---  {\it the AGB matter contains higher CNO than the pristine first generation matter}. As the 3DU also brings to the surface products of s--process nucleosynthesis, the second generation stars formed at {\sl late} times will be s--process enhanced \citep[first with the products of the $^{22}$Ne neutron source, than with the products of the $^{13}$C neutron source, e.g.][]{busso1999ARAA}.\\
At a \Thbb\ very dependent on the metallicity, but approximately corresponding to masses $\sim$3\msun, the HBB process is no longer efficient, and the 3DU transforms the AGB into a carbon stars. As Carbon rich matter has never been found in second generation spectra, we must conclude that, if AGBs are the polluters, the entire process of multiple populations ends before, or well before, the 3\msun\ evolve ---in our models, at ages $<$300Myr.\\

\subsection{Dilution as a necessary input}
\begin{figure}
\vskip -35pt
\centering{
\includegraphics[width=8.5cm]{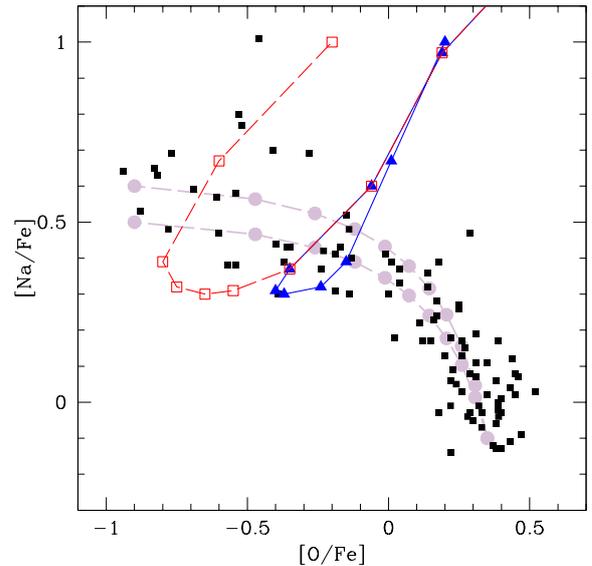}
\vskip -65pt
\caption{Black squares represent the O--Na data for the giants of NGC~2808 from Carretta et al. 2009. Superimposed is the O--Na relation from the yields in Ventura et al. 2013 for Z=0.001 (blue line with triangles, each triangle represents the yield of a different mass) and its modification by assuming that the whole super--AGB range progeny suffers deep mixing which depletes oxygen by a further factor (red open squares and dashed red line). The masses for which oxygen is reduced are (starting at the top right end of the red sequence) 8, 7.5, 7, 6.5, 6.3, 6 \msun. For masses 5.5, 5 and 4.5 \msun the yields are not changed (red squares and blue triangles coincide).  Two semi empirical dilution curves are shown as grey lines with dots (see text). }
}
\label{figure:fig2}
\vskip -10pt
\end{figure}

A well known problem of the abundances of O and Na in AGB ejecta is that, when examined as a function of the initial mass, they provide an O--Na {\it correlation} instead of the well known {\it anticorrelation} which is a typical signature of multiple populations. On the other hand, the distribution of O and Na data, especially seen in the clusters which show an extended O--Na anticorrelation, resembles a ``dilution" curve. In fact, assuming that the ejecta composition corresponds to the extreme abundances of O and Na in the observations, while the standard O and Na of halo stars is the ``pristine gas" composition, the intermediate abundances can be the result of star formation out of pristine gas and a  decreasing amount of ejecta \citep[see, e.g.][]{dantonaventura2007}\footnote{A dilution model is also used in the FRMS approach, e.g. to account for the Na--Li observations in NGC~6752 \citep{decressin2007}. Formally, the massive stars O--Na yields are anti-correlated for decreasing initial mass, but their anticorrelation does not match the observed one \citep{bastian2015}. }. In Fig.~\ref{figure:fig2}, superimposed to the O--Na data for the giants of NGC~2808 \citep{carretta2009a},  we show  two semi empirical dilution curves (in grey). The abundances along these lines vary from 100\% AGB matter, at [Na/Fe]=0.5 or 0.6, and [O/Fe]=--0.9, to 0\% AGB matter (pure pristine gas composition) at  [Na/Fe]=--0.1, [O/Fe]=+0.35. The grey dots, from left to right, represent dilution from 0\% pristine gas to 10, 20 ... 100\% of pristine gas. It seems unlikely that specific, different AGB models may produce a O--Na anticorrelation that resembles the dilution curves. \\
Dilution is also required to explain the Lithium versus  Sodium anti correlation in  NGC~6397 (D'Antona et al. 2011, and references therein).

The requirement of dilution between ejecta and pristine gas opens a number of interesting and complex questions concerning the gas dynamics during the cluster formation phase and the details of the formation and early dynamical evolution allowing the necessary mixing of pristine gas with AGB ejecta \citep[see e.g.][]{trenti2015}.
Here, however,  as discussed in the Introduction, our attention is focussed on the chemistry of SG populations which, at present, provides the strongest observational constraints on the possible sources of polluted material.

It is important to point out that efforts aimed at
assessing the viability of the AGB models with dilutions are often based on oversimplified assumptions. 
Key fingerprints of multiple populations, such as the O--Na anticorrelation, have been often modeled by assuming that all the polluting gas has a unique chemical composition, and that the different abundances are due to different ``dilution" with pristine gas characterized by the original (first generation) composition \citep[e.g.][]{conroy2012}. This hypothesis can be reasonable only in two cases: 1) if the formation timescale of the second generation is very short, so that the ejecta composition have no time to appreciably change; 2) if the AGB gas accumulates in the cluster for a long time and is well mixed, until re--accretion of pristine matter induces a unique star formation (SF) episode, with stars forming with different degrees of dilution between the AGB ejecta and the accreting gas.\\
The widespread presence of discrete populations which is likely to be due to multiple bursts of star formation is a strong indication that {\sl not only different dilutions, but also different AGB ejecta compositions are at play}. Thus, for example, the observational result that a unique dilution curve can not explain the Al--Mg data in NGC~6752 \citep{carretta2012} and  NGC~2808 \citep{carretta2014}, provides further support to this hypothesis.

\section{Multiple generations: a sequence in time}
\label{sec:4}
Summarizing the result of Fig.~\ref{ff1}, multiple populations in the AGB scenario are the result of SF in matter contaminated by proton capture nucleosynthesis, in a variety of physical conditions, which may include ---or not--- some light nuclei. We may think that the SF proceeds for several tens of Myrs, although it must definitely halt before the AGB ejecta become richer in C than in O. \cite{dercole2008} proposed and showed that the end of SF may be linked to the onset of SN\,Ia explosions, but other mechanisms may be considered too.  

In  \cite{dercole2008}  SF proceeds in a continuous way, as the conditions for SF are met in the cooling flow, in absence of perturbing events. Looking at the clustering of stars in the \cite{milone2015} CNO--two--color plane, or at the distribution of abundances in \cite{carretta2015}, it seems more likely that different populations are formed in separate bursts of formation \citep{dercole2012},  following triggering events which remain to be investigated. 
Separate bursts may also be thought of as an indeed continuous SF, going on until some perturbing event stops it for a while, and then SF resumes again with different modalities (e.g. the degree of dilution). \\ 
The duration and modalities of the formation events will vary from cluster to cluster. In NGC~2808 the process must have lasted long enough to lead to the formation of at least 5  populations differing in chemical abundances.

We have already proposed some models able to reproduce some of the key properties of multiple populations in NGC~2808 \citep{dercole2008, dercole2010, dercole2012}. More recent observations confirm the presence of the three main helium groups, which in \cite{milone2015} are identified as the first generation (B), the ``extreme" population born from pure AGB ejecta (E), followed in time by an ``intermediate"--helium population (D), in which re--accretion of pristine gas plays a role. 

Is it possible to extend the model in order to explain populations C and A? We show that this is qualitatively possible. In the following, a full chemical and dynamical model, which is beyond the scope of this paper, will be necessary to reconstruct the whole formation history of NGC 2808, along with the initial mass and structural properties necessary to reproduce all its current observed properties.

\subsection{A note on the oxygen yields}
\label{oxygen}
The discovery that NGC\,2808 hosts three main sequences  \citep{dantona2005,piotto2007}, which could only be interpreted in terms of difference in the initial helium content in the forming gas, was one of the important drivers to interpret the chemical anomalies of globular clusters stars in terms of different stellar generations. \cite{milone2012mf2808} estimate a helium difference $\Delta$Y$\simeq$0.13 between the red and the blue main sequence, using standard stellar models not including any iron variation. This provides Y$\sim$0.38, if the red main sequence is located at a standard Y$\sim$0.25.
The helium difference between the main sequences A and E is estimated again to be $\Delta$Y$\simeq$0.13 in \cite{milone2015} ---their Table 3--- but the first generation is now identified with the B population, and $\Delta$Y is $\sim$0.10 between E and B (Table\,\ref{tablemilone}, lower panel), giving Y$\sim$0.35 for the blue main sequence. The AGB models maximum helium yield is Y$\sim$0.36 in our models (see Fig.\,3) very similar to the result by \cite{doherty2014}. Values up to Y=0.375 are found in the super--AGB models by \cite{siess2010}. Therefore, {\it if the E population forms from the most massive pure AGB ejecta, as we will assume,} there is good agreement between observations and models. 

On the other hand, the \cite{ventura2013} models have a difficulty in reproducing the most extreme oxygen abundances. This is a long standing problem\footnote{We again have to emphasize that massive star yields {\sl do not} provide a better fit to the oxygen extreme abundances. The matter ejected from the FRMS can be highly depleted in oxygen in the inner stellar layers processed by core H--burning. But, when formation occurs in the `excretion' disk around the star \citep{chantereau2015}, the helium abundance in these stars will be much larger (up to Y=0.8) than the values which are attributed to the extreme stars from the main sequence split. There is no observational evidence for the presence of such stars, and the helium richest main sequences are clustered, with a small scatter, below Y$\sim$0.40.}. 
When super--AGB models were not yet available, \cite{dercole2010} assumed semi empirical yields with low O in their first modelization. But the yields of super--AGB stars resulted in oxygen abundance larger than those of the most massive AGB \citep{ventura2011sagb}, so \cite{dercole2012} proposed that the very low oxygen abundances were due to deep mixing in the giants of the extreme population, favored by their large helium abundance \citep{dantonaventura2007}. In this work, we make the same assumption but we point out that, as discussed in \S \ref{sec:agbnuc}, alternative solutions are possible: a larger oxygen depletion may be achieved by reducing the mass loss rate in the models \citep{ventura2011}, a reduction which must go together with a reduction of the  cross section $^{23}$Na(p,$\alpha$)$^{20}$Ne.

Fig.\,2  shows the O and Na yields computed by \cite{ventura2013}, and the different oxygen values assumed in this work for M$\geq$6\msun.

\subsection{Outline of the BEDCA model}
\begin{figure*}
\vskip -10pt
\centering{
\includegraphics[width=6cm]{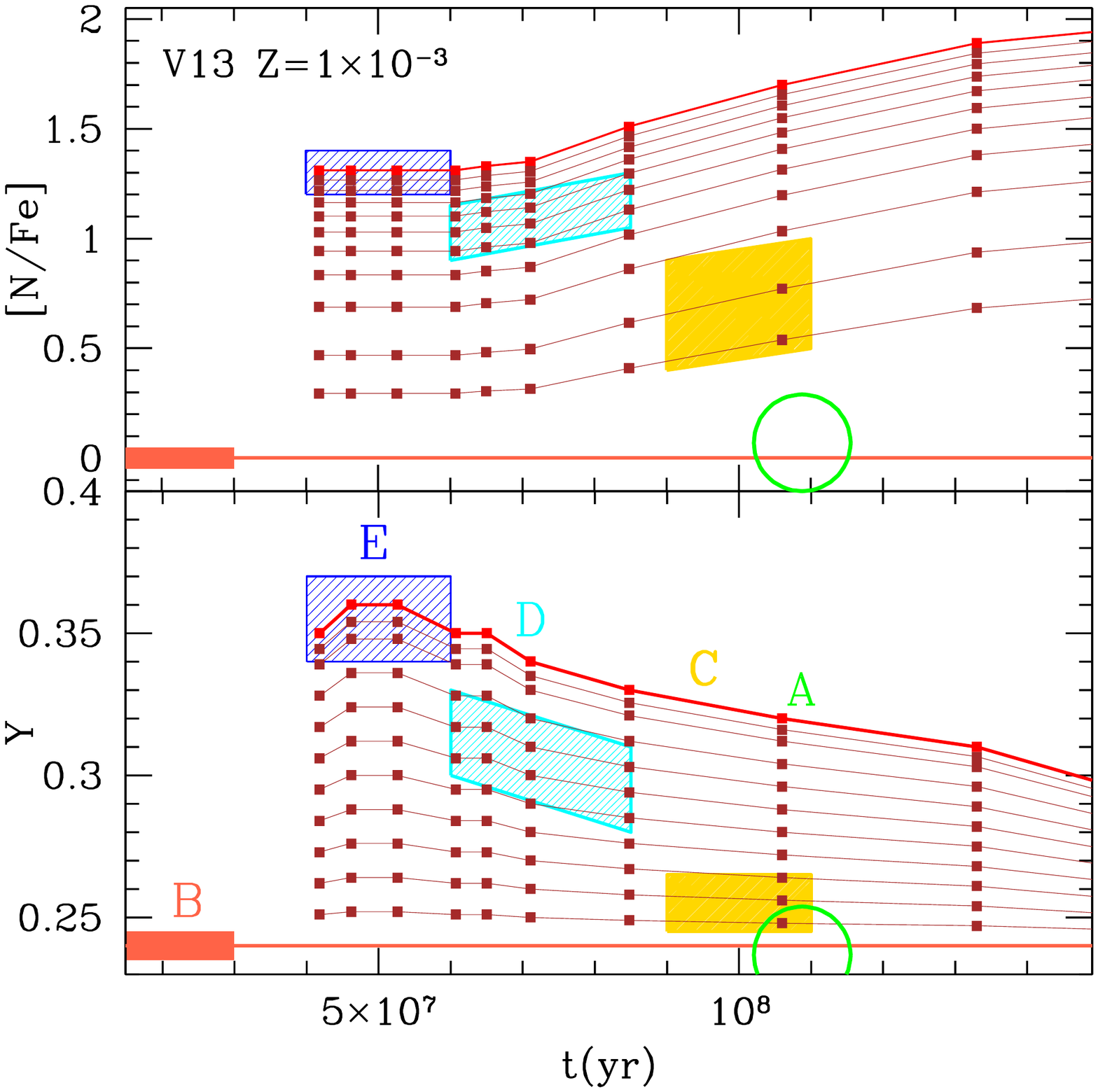}\includegraphics[width=6cm]{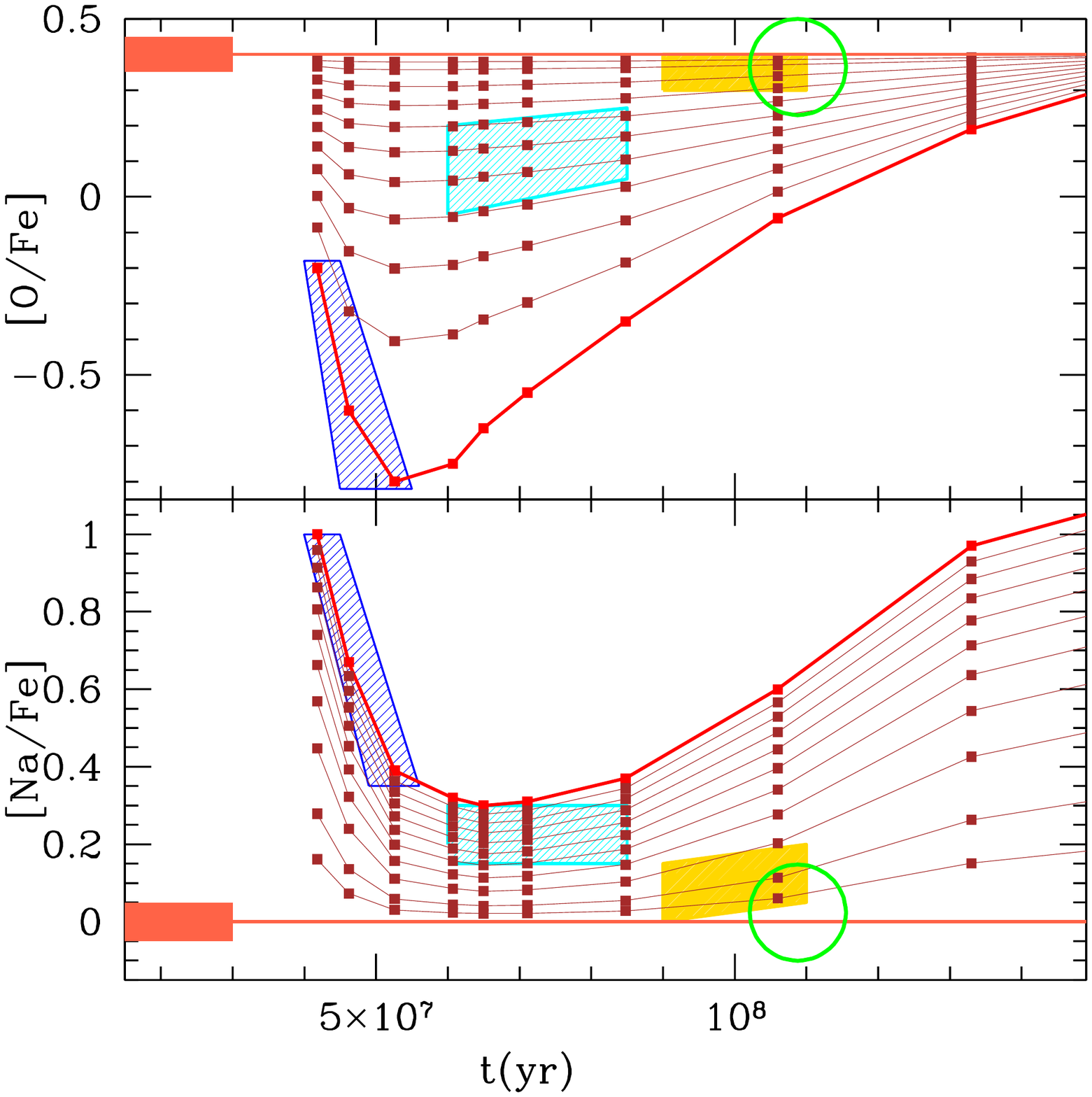}\includegraphics[width=6cm]{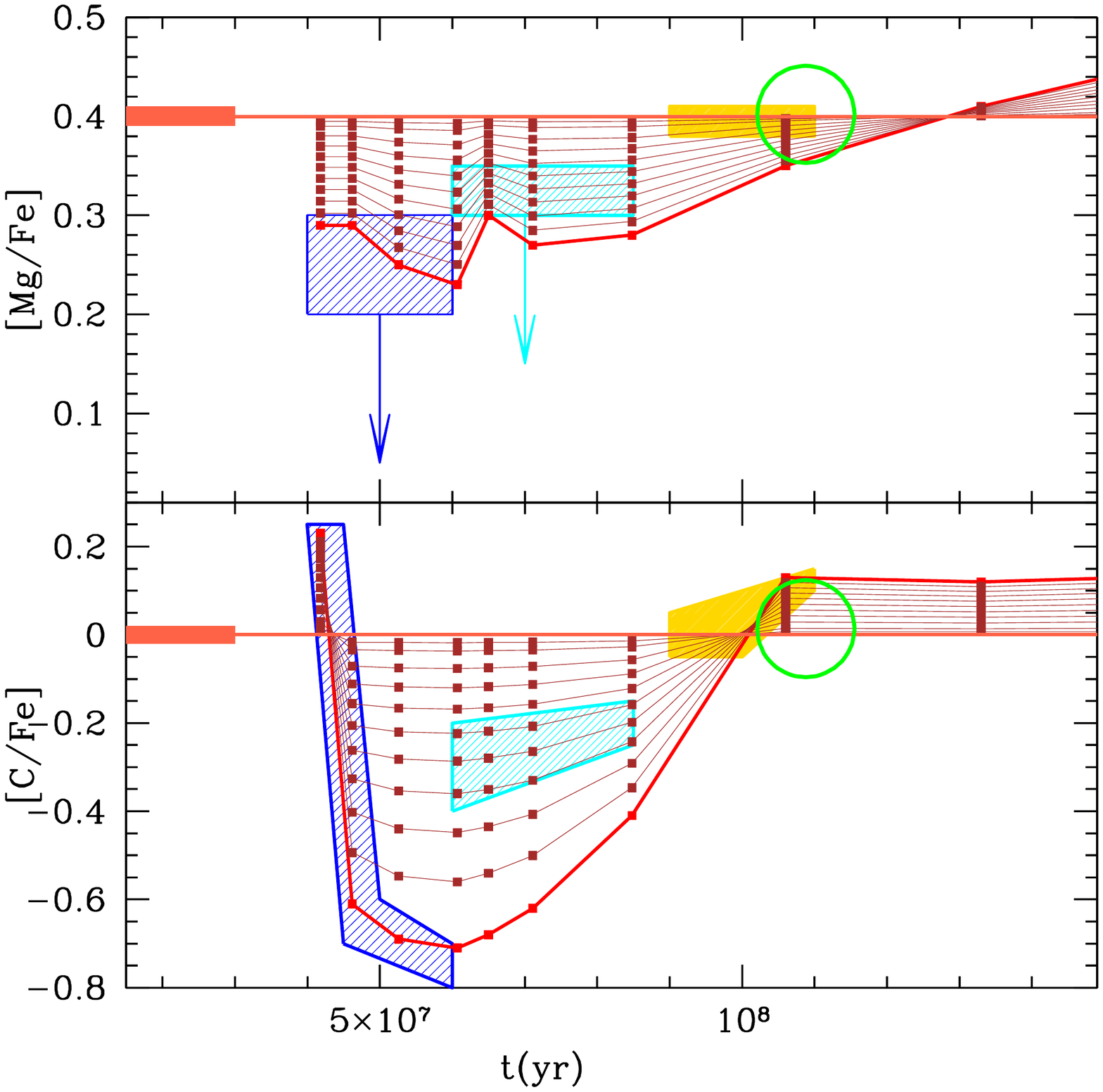}
}
\vskip -55pt
\caption{Each panel shows the average abundance of nitrogen (top left), helium (bottom left), oxygen (top center), sodium (bottom center),  magnesium (top right) and carbon (bottom right) versus time in the ejecta of massive AGB stars having Z=10$^{-3}$ (Ventura et al. 2013). The yield for each initial mass (8, 7.5, 7, 6.5, 6.3, 6.0, 5.5, 5.0 4.5 \Msun, from left to right) is shown by dots. The abundances of oxygen at t$< 7 \times 10^7$yr have been adjusted as explained in the text. The pure ejecta abundances are shown as a red line with squares. The initial abundances in the AGB stars is represented by the horizontal red lines without dots. Abundances in diluted matter are shown by the brown lines with squares. Starting from the initial FG abundance lines, the AGB gas percentage  in each curve is 0.05, 0.1, 0.2, 0.3... 0.9. 
In these planes, we show our guess for the location of the five populations defined by Milone et al. (2015) for the cluster NGC~2808: 1) B (red) is the first generation population, having the initial gas abundances (red rectangles, stars born at t=0); 2) E is the most extreme second generation, born from pure AGB ejecta, defined by the blue rectangles; 3) D is the `ìntermediate" second generation, in which the AGB ejecta constitute 40-60\% of the forming gas  (cyan regions); 4) C, the new population discovered in Milone et al. 2015, with normal helium and high nitrogen: this is obtained by very strong dilution of the AGB ejecta which already have a strong effect of 3DU, with CNO enhancement; 5) A, the small population at the red of all other sequences, is here identified with a remnant star formation, mostly from first generation gas contaminated by the ejecta of one SN\,Ia (green circle).
The Mg boxes for population E and D (blue and cyan) in the bottom right panel show with arrows the observed location in the data by Carretta (2015), $\sim 0.15$dex lower than the models. As explained in the text, a lower mass loss rate would provide the correct result, but it would reduce the sodium abundances. }
\label{figure:fig3} 
\end{figure*}

The \cite{milone2015} and \cite{carretta2015} data imply that star formation in the cluster occurs in separate bursts. 
We use the main observational constraints to build a very simple model and outline the possible epochs during which each of the populations can be born, and the corresponding average values of dilution necessary to reproduce the available abundance patterns (Table\,\ref{tablemilone}). As we have discussed, the yields of the models we are using will definitely require further revisions. Anyway,  the fundamental properties of the yields, which are linked to the evolution with time of \thbb,  emerge clearly, as we are going to show.\\
The first part of this interpretation was outlined and defined with more complete models by \cite{dercole2008, dercole2010, dercole2012}, and we start from what emerged from those investigations. The novelty is the possible interpretation of the two new populations (C and A) identified by \cite{milone2015}, in the context of that same model.
The key ingredients of our results are summarized in Fig.~\ref{figure:fig3}.
The starting abundances of N, He, O, Na, C and Mg in the models by  \cite{ventura2013} are shown as horizontal lines. 
We also plot  the abundances of the same elements in the yields of different masses, for Z=0.001, [$\alpha$/Fe]=0.4 \footnote{The Oxygen yield has been modified according to Fig.~2.} The initial abundances and the yields delimit the region of abundances in `diluted' gas.
Starting from the pure yield lines, we plot the lines for mixtures in which the AGB gas constitutes a fraction of 90, 80, 70, 60, 50, 40, 30, 20, 10 and 5\% . On these schemes we build our interpretation of the BEDCA sequence summarized below, and explained better in the following sections.

{\bf B:} both the chemical abundances of all light elements \citep{carretta2015} and the location in the  CNO--two--color diagram show that these are the first generation stars. Thus we represent it by red boxes at the left of the AGB epoch (t$<$30Myr), placed on the initial abundances.

{\bf E}: this group has been early recognized as the one in which nuclear processing has been most extreme, with large depletion of Oxygen, processing of Mg, production of Si and of K. Thus in AGB modeling this requires the largest \Thbb's, and we place it at ages 40--60Myr. \cite{dercole2008} first suggested to model the blue main sequence of NGC~2808 as due to SF in the pure ejecta of the most massive AGBs, in which Y is indeed very similar and very large \citep[Y=0.35--0.38 in all published stellar models ---][]{siess2010,ventura2013,doherty2014}. Chemical modeling by \cite{dercole2010,dercole2012} showed that also the other chemical abundances could be reproduced under this hypothesis. Thus we schematically represent group E with the blue boxes, lying on the AGB yield line. The masses involved are the super--AGB masses from 6--6.5 to 8\Msun. \\
We emphasize again that these values of mass and ages apply to the models we use here, the ranges may be different for other models, the main characteristic of this population is that it is born from the ejecta of the first (most massive) AGBs born after the end of the SN~II era. 
The peculiar carbon yield of super--AGBs shown in Fig.~3, right bottom panel, is discussed in \S~\ref{CNO}. 

{\bf D}: we propose that the discontinuity among group E and D is due to a sudden onset of dilution with pristine gas, at an epoch $\sim$60\,Myr after the cluster formation \citep{dercole2010, dercole2012}. Fig.~\ref{figure:fig3}  suggests star formation in a mixture in which the AGB ejecta constitute 40--70\%, and extending for $\sim$25~Myr  (cyan boxes). This again is in line with our previous chemical evolution models.

{\bf C}: a discontinuity in the star formation must occur at a time 85--90\,Myr, followed by an additional star formation event in gas which is scarcely contaminated by the AGB ejecta (5--15\%). A time of formation from about t=90 to about t=110Myr is suggested in the figure, and is marked by the yellow boxes. With such a huge dilution, the abundances go down to values very similar to the first generation values. Y may be larger than the initial value by $\sim$0.01--0.02 (an important hint, which we discuss in \S~\ref{HB}). Na is in the range from pristine to $\delta$Na=+0.2dex, as observed, but, most importantly, {\it Nitrogen may be larger than the pristine abundance by 0.4--0.7\,dex, as predicted the differential analysis between the spectro-photometry of groups B and C by \cite{milone2015}.} 
We emphasize that such a population can be in part hidden and overlap with the first-generation in the Na-O plane \citep[group Primordial 2 by][]{carretta2015} and can be fully identified as a separate second-generation population only when studied with spectro-photometric or spectroscopic observations sensitive to the N abundance.

{\bf A:} We suggest that the population A is formed towards the end of the C group formation epoch, from mainly pristine gas polluted by the ejecta of the first SN\,Ia exploding in the cluster. 
We are not guided by spectroscopic data, but by considerating the  following.
Although the `A' MS data are not well separated in the CNO--two--color diagram, \cite{milone2015} show that it may be slightly redder than the B MS, and suggest that these are stars of the first generation affected by a slight iron increase ($\delta$[Fe/H] $\sim 0.1$\,dex) attributed to type II supernovae (but see also \S\,\ref{sec:obs}). Here we revise their interpretation (see \S\,\ref{sec:Agroup}).
We identify group A in Fig.~\ref{figure:fig3} with green circles, whose location in time and composition is of course very arbitrary. Actually, their formation could be contemporary  to group C formation, in a fraction of gas enriched by the first SN\,Ia. If A stars are the last one to be born in the cluster, they too {\it might have some N enhancement}, and this could be the reason for their location in the CNO--two--color diagram. Spectroscopic observations are necessary for this group.

\begin{figure}
\vskip -40pt
\centering{
\includegraphics[width=8.5cm]{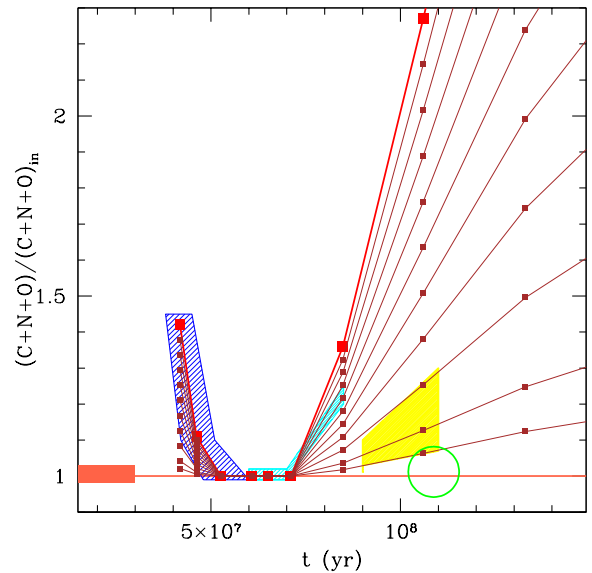}
\vskip -55pt
\caption{abundances of CNO in the ejecta. The figure adopts the same display as Fig.~3 }
\label{figure:4}
}
 \end{figure}
 
 \subsection{Abundances of other light elements}
 \label{sec:others}
 The abundances of Carbon are shown in the bottom right panel of Fig.~3. Carbon is always very depleted in the ejecta, apart from the most massive super--AGBs, where it reaches $\sim$0.2\,dex. Several computations show that,
 at low--metallicity, the second dregde--up occurs in the ``dredge--out" modality \citep{ibenritossa1997, siess2006, siesspumo2006}. During the inwards penetration of the base of the external envelope, a convective zone first forms within the helium--burning, then, after growing in mass, it merges with the surface convective region. This process favours a significant increase in the surface carbon. This kind of  event is much less studied than HBB or the 3DU, and so it must be considered uncertain. Nevertheless, the peculiar dredge--out mechanism occurs only in the most massive models, and we can expect an enhanced C--abundance in the (small) fraction of the population E born from these ejecta. If this is not found, it is possible that these stars evolve into e--capture supernovae \citep[in this context, see the discussion in ][]{pumo2008}.
 
We show in Fig.~3 the abundances of magnesium (top right panel). The agreement with the spectroscopic result is only qualitative here. As we mentioned in \S~\ref{sec:agbnuc},  \cite{carretta2015} has shown that also the Mg abundances in NGC~2808 indicate the presence of  5 groups. Quantitatively, our models fail to reproduce the Mg depletion of 0.4~dex shown by these data for the most extreme stars (Figure 8 in the Carretta paper, and Table\,\ref{tablemilone}).The maximum depletion of the models is reached for the mass 6\,\msun, and amounts to 0.17\,dex, which would mean $\le$0.1\,dex if we operate a dilution by $\sim$50\% with pristine gas. We discussed in \S~\ref{thbb-time} that this problem is due to the mass loss rates of the models, and these, in turn, depend on the necessity of preserving sodium. 
Therefore in Fig.~3 we add arrows to the boxes, which represent the groups E and D locations from the models. The end of the arrows indicate where the boxes should be placed to be consistent with the data.

We have not discussed specifically aluminum, which the \cite{carretta2015} data show to cover a wide range of abundances $\sim$1.1\,dex. Also our models display an abundance range of a factor $\sim$10. A discrepancy with the data is present only for group E, representing SF in the ejecta of the most massive super--AGBs and AGBs. For super--AGBs  the models provide only $\delta$Al=+0.4--0.9\,dex, another problem which is probably linked to the high and unconstrained mass loss rates of the models.

Potassium looks bimodal  in the \cite{mucciarelli2015} data, and only the E population has $\delta$K as large as $\simeq$+0.2. This is consistent with the high \thbb\ of the super--AGB models, but again we need smaller mass loss rates to achieve this result \citep{ventura2012ngc2419}.

\subsection{The CNO content of population C}
\label{CNO}
Fig.\,\ref{figure:4} is complementary to Fig.\,3 and displays the time evolution of the C+N+O yields in the \cite{ventura2013} data, and their dilution curves, as a function of time. The  C+N+O increase at small ages is due to the peculiar ``dredge--out" of Carbon in the most massive super--AGB models, previously discussed in \S~\ref{sec:others}. The increase in CNO at ages $>$80~Myr is an effect of the 3DU, already discussed in \S~\ref{thbb-time}. 
Although CNO increases rapidly with decreasing mass,  C is still CN processed in the envelope, so {\it the dominant increase is that of the N abundance}, which we have seen in Fig.\,3. 
We claim that the N increase due to the 3DU is allowing us to identify stars belonging to the `C' group, and separate them from the first generation stars of group B. 

Notice that the processing of N inside the helium inter-shell (at each episode of dredge up) has two effects: 

1) provide the neutrons for s--process from the chain \\
$^{14}$N($\alpha,\gamma)^{18}$F($\beta,\nu)^{18}$O($\alpha,\gamma)^{22}$Ne($\alpha$,n)$^{25}$Mg\\
so that also s--process enhancement begins in the envelope; 

2) the primary $^{22}$Ne dredged up will capture protons and further increase Na in the envelope. The sodium increase may be seen in the C group, in spite of the huge dilution of the ejecta. On the contrary, we do not expect that the s--process enhancement is seen with such a huge dilution. 

In our schematic model, and taking at face values the uncertain CNO increase of \cite{ventura2013} models, the total CNO in group C should be a factor 1.1--1.3 larger than the initial value. 

\subsection{What is the population A?}
\label{sec:Agroup}
 At an age above $\sim$100~Myr, SN\,Ia will begin to explode in the cluster. 
A single SN\,Ia ejects $\sim$0.8\msun\ of iron  \citep{nomoto1984, thielemann1986}.  If it were redistributed uniformly, this iron mass would be enough to change by $\sim$0.1\,dex the [Fe/H] of 50000\msun (about 6\% of the cluster present mass) starting from the initial iron content corresponding to a metallicity Z=0.002.
The {\it first} SN~Ia exploding in the cluster may not able to stop the cooling flow, if it is not followed soon by  other supernova explosions \citep{dercole2008}, so the `A' stars may in fact have been formed by matter polluted by this first SN\,Ia explosion, just prior to the end of the multiple population epoch. After this last episode of star formation, SN\,Ia begin to explode close enough in time that the star formation in the cluster is definitely inhibited. Of course, there may be a variety of different situations in other clusters, and we further discuss this issue in \S~\ref{sec:sn1a}.

\section{The Horizontal Branch morphology in NGC~2808}
\label{HB}
\subsection{The `C' group location in the HB}
The abundance analysis of the horizontal branch stars had already pointed out the existence of population C.\\
The B (primordial) and C (late second generation) groups can be identified with the two peaks in the [O/Na] spectroscopic distribution by  \cite{carretta2015}, at [O/Na]$>$0. Carretta shows that the high [O/Na] stars are divided into two subcomponents, with small difference in [O/Fe] and larger differences in Na. He also correlates the presence of these two groups with the O--Na distribution of stars in the red side of the horizontal branch (RHB) of the cluster. \\
In the first attempts to model the HB of NGC\,2808 with multiple populations  \citep{dc2004}, it was proposed that the RHB should contain only standard helium, first generation, stars, and that the blue side of the HB should be populated by He--richer stars of second generation, down to the extreme HB. The hottest group  would contain very He--rich stars progeny of the blue main sequence \citep{dantona2005, dc2008}, a result also confirmed by \cite{dalessandro2011}. \\
Later on,  \cite{gratton2011hb2808} found that an O--Na anti correlation (although not fully extended) is instead present also among the RHB stars. On this basis, Gratton et al. were the first to suggest that more than three populations were present in the cluster. \\
A subsequent spectroscopic analysis of RHB stars \citep{marino2014hb2808} showed a more marked dichotomy in Na, and that the group richer in Na appeared to be slightly more luminous and bluer than the red group. Also this work attributes this feature to the presence of another population hidden among the RHB stars, and explicitly relates it to a group in which the AGB ejecta have suffered more dilution\footnote{``Following the D'Ercole et al. scenario, we suppose that they are the latest stars to have been formed from highly diluted material, such that their abundances in light elements and Y approach the primordial values of the first generation. Hence, these stars could have formed after the intermediate second-generation stars, which show evidence for a higher degree of AGB pollution. If this prediction is correct, the abundance pattern of light elements of these stars is dominated by dilution with pristine gas, and the red MS and the RHB contain the first and last stars formed in the cluster." \citep{marino2014hb2808}}. Now finally, the scenario proposed finds a confirmation from the CNO--two--colors diagram.

Nevertheless, something remains to be understood.  This highly--diluted population C may have a slightly larger helium mass fraction, larger by about 0.01--0.02 than the first generation abundance. According to \cite{dc2004}, this small variation should be able to shift the HB stars at least into the RR\,Lyr gap, or even at its blue side, due both to the high dependence of the \teff\ of models on the mass, for NGC~2808 metallicity and for this range of \teff, and to the decrease of the evolving mass in the RGB on the initial helium content (see Fig.~5).
The RR\,Lyr gap in this cluster is actually deprived of stars, with respect to the red and blue side, and this feature has been interpreted  as the result of a helium discontinuity between the RHB and the reddest HB stars at the blue side of the gap. This apparent difficulty turns out to be another point lending further support to our model for the C population.

\subsection{The HB stars distribution: a function of helium and of C+N+O enhancement}

In NGC~2808 stars we have  [Fe/H]=--1.18$\pm$0.04, and $[\alpha$/Fe]$\sim 0.2$\ \citep{carretta2009c}, so the global metallicity must be in between 1 and 2$\times$10$^{-3}$. We use Z=2$\times$10$^{-3}$\ to discuss the HB properties.
We show in Fig.~5 a representation of the classic explanation of the HB morphology in this cluster. 
The four lines in the lower part of the figure are the mass versus \teff\ relations for HB zero age horizontal branch (ZAHB) models of Z=2$\times$10$^{-3}$ and Y=0.25 (full line, black) and Y=0.28 (lower full line, red), [$\alpha$/Fe]=0.4 for standard CNO abundance ratios. The two dot--dashed lines are the mass--\teff\ for models having Y=0.25 (blue) and 0.28 (magenta) in which the CNO total abundances has been increased by a factor 1.5. The \teff\ boundaries of the RR~Lyr gap are represented as the two vertical (dashed, cyan) lines. The location of the red HB stars analyzed by \cite{gratton2011hb2808} and \cite{marino2014hb2808} is represented by the green arrow.\\
We plot at the top the evolving RGB mass at 12~Gyr for Y=0.25, 0.26, 0.27 and 0.28. The CNO enhanced masses are about the same as the CNO--standard ones. We assume that RGB stars having  Y=0.25  loses the right amount of mass to put it on the RHB (0.224\msun\ for the case shown). 
From the flatness of the mass--\teff\ relation towards larger \teff, we can appreciate that the mass loss spread for the population at standard Y must be kept within $\sim$0.01\msun\, so that the location does not fall into the RR\,Lyr region. As an example, the color width of the red clump is fit with a mass loss having gaussian standard deviation $\sigma$=0.008\msun\ in \cite{dc2008}.
For each Y, and assuming the same mass loss, we mark as squares the intersection with the corresponding mass--\teff\ line. The scheme shows that an increase in Y shifts the models to larger \teff, and that the scarcely populated RR~Lyr gap in NGC~2808 \citep{clement1989rr2808} requires a jump in Y by $\sim$0.01 to be modeled \citep{dc2004}. Thus, if  the C group stars had only a slightly enhanced  Y=0.26 or even a bit larger, they should fall at the left of the RR~Lyr region. 
On the contrary, and increase in CNO for the same Y=0.25 shifts the model to cooler location (blue dot), contrary to the observations. A small increase in Y in CNO rich models brings back the model to the \teff\ range of B red clump stars (orange dot).

\begin{figure}
\vskip -35pt
\centering{
\includegraphics[width=8.5cm]{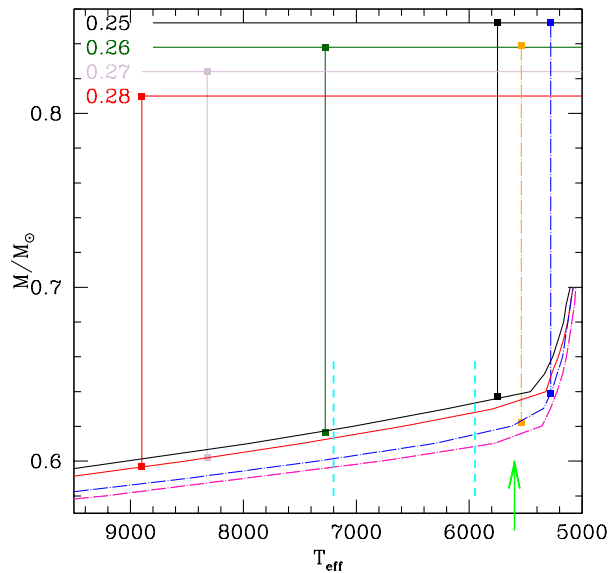} 
\vskip -55pt
\caption{In the lower part of the figure, the mass--\teff\ relations for ZAHB models of Z=0.002 are shown for Y=0.25 (top line, black) and Y=0.28 (red line) for standard C,N and O abundances. The two dash--dotted lines below represent the mass--\teff\ location of models in which the total CNO is 1.5 the standard CNO, for Y=0.25 (upper line, blue) and Y=0.28 (lower line, magenta). At the top, the four horizontal lines  mark the value of mass in evolution in the RGB  at age of 12~Gyr, with different Y, as labelled (the evolving mass does not change for different total CNO). The vertical segments represent a constant mass loss of 0.22\msun, and thus the location along the ZAHB obtained for different Y. At constant CNO, even a very small helium increase by 0.01 shifts the location from the red clump to beyond the RR~Lyr gap (cyan vertical dashed lines). 
A similar location in \teff\ of B and C groups in the red clump \citep[green arrow, from][]{marino2014hb2808}, as shown by the two black and orange dots, is possible if the C group has both a larger CNO and a slightly larger Y (for the same mass loss in RGB).
}}
\label{figure:fig5} \end{figure}

\section{Cluster to cluster differences expected at different metallicity}
\label{sec:metals}
Before we discuss the general case of multiple population formation, we make few simple considerations on the literature data which already show that an analysis based on AGB ejecta pollution satisfies some basic observations.
Fig.~1 shows that the maximum \thbb\ reached in AGBs depend on the metallicity of the models. This is a first order result, which basically depends mainly on the opacities in the envelope: more transparent envelopes (lower Z) allow larger \thbb. A very simple consequence is that {\it higher metallicity clusters can not show the same degree of p--capture processing as the lower metallicity clusters}. Of course, there may be low metallicity clusters which {\it do not} show signs of the presence of extreme populations, this depends on the perturbing events, discussed in \S~\ref{frame} which modulate the SF history. But, for similar conditions in the SF timescale, that is, when similar masses contribute to the ejecta, we do not expect to find the same extreme compositions in second generation of clusters of higher Z.

We list a few results lending support to this point.
\begin{enumerate}
\item  potassium, which requires \thbb$>$125MK to be formed by proton captures on argon \citep{ventura2012ngc2419} is found to vary only in NGC~2419 \citep{cohenkirby2012} and in NGC~2808 \citep[][]{mucciarelli2015}. Notice that both clusters have an extreme population \citep[for NGC~2419, see ][]{dicrisci2015}, which we interpret as born from pure ejecta, and thus are best qualified to show such p--capture extreme product.  None of other examined clusters show variations \citep{carretta2013}, in particular, no hint of variation is present in 47\,Tuc.
\item In 47\,Tuc ([Fe/H]$\sim$\,--0.7) the maximum magnesium processing is by $\delta$Mg=\,--0.1 \citep{thygesen2014}
\item Aluminum varies by a decade in low and intermediate metallicity clusters, but it looks like its variations are much smaller at larger metallicity, see the discussion in \cite{cordero2015}.
\end{enumerate}

\section{The general framework}
\label{frame}
We conclude that the populations in NGC~2808 were formed in the order of the acronym BEDCA, and that their characteristics are consistent with the main features of each of these populations, including a good explanation for the group C, and a working hypothesis for the role of SN\,Ia in the group A.
Thus the multiple populations in GCs are recognized to be the result of SF events which may be separated by the occurrence of different triggering events, allowing transitions in the SF resulting in chemical differences between the groups. The \thbb\ reached by the models are also large enough to allow the contemporary processing of magnesium, and formation of aluminum and silicon, as found in the most extreme populations of NGC~2808. Although the yields of the models adopted do not reproduce the quantitative depletion in the abundance of Mg, or the entire production of Si,  we have discussed how this problem can be solved thanks to a smaller mass loss rate. This can be assumed only if the  cross section $^{23}$Na(p,$\alpha)^{20}$Ne is smaller, so that sodium can be preserved, at high \thbb, also with a longer HBB evolution.  

It is clear that we can further generalize the scheme shown in Figs. 3 and 4, by further subdividing the star formation events, as probably required by the finer discreteness of the data. For example, we schematically represented five different boxes of chemical properties for the five groups defined in \cite{milone2015} and attributed the separation between the groups to triggering events (onset of dilution, delayed SN~II supernova explosions). As already mentioned, within these boxes other minor separations can be present, due to other triggers, so the model can be refined once we know better the amount of discreteness in the data. 

We now address two more general questions: 

1) what are the possible events which provide the transition between the different star formation epochs? 

2) How can the NGC~2808 scheme be adapted to predict the behavior of the chemical differences among multiple populations in other clusters?

\begin{figure}
\centering{
\includegraphics[width=8.5cm]{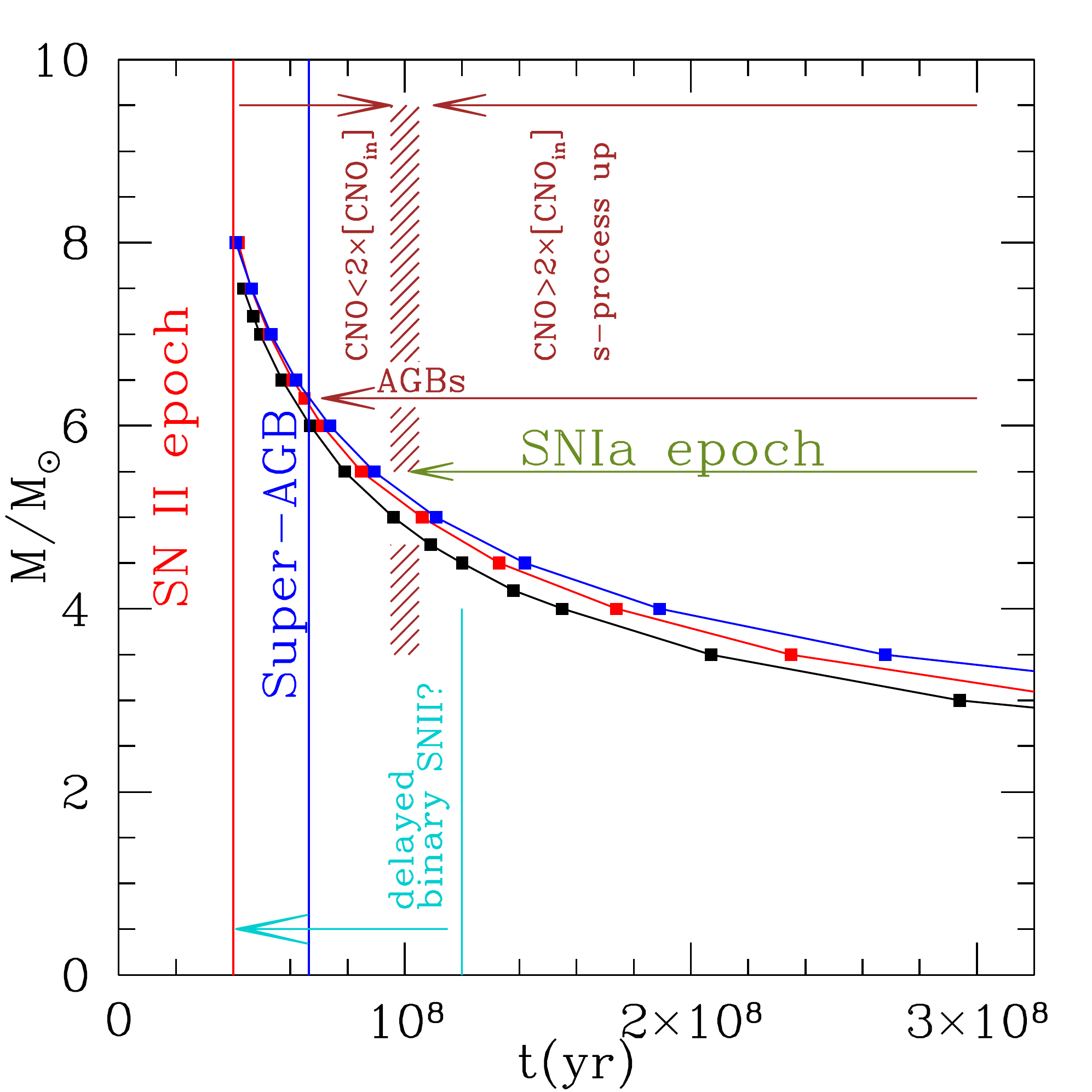}
}
\caption{ The mass evolving as a function of the age is shown, from Ventura et al. 2013. Times are dependent on the assumptions made in the models, especially on the core overshooting during the core H--burning phase. The end of the supernovae Type II epoch in these models is at $\sim$40~Myr, and is followed by the epoch of super--AGB evolution which lasts $\sim$20~Myr. Delayed type II supernovae may explode until the evolutionary age of the primary mass which may provoke the explosion of the companion, by increasing its mass beyond the single--SNII minimum mass. The age at which SN\,Ia begin to explode is uncertain, but it  is dependent on the age at the end of the super--AGB phase, which tags the beginning of formation of C--O white dwarfs.}
\label{figure:fig6} 
\end{figure}

The specificity of these patterns depends primarily on the metallicity of the cluster stars.
The presence of separate bursts depends on other possible parameters, intertwined with each other. We consider specifically: 
 \begin{itemize}
\item the timing and extent of dilution of nuclearly processed AGB ejecta with pristine gas;
\item the role of binary SN\,II supernovae in delaying or halting the formation of the second generation;
\item the role of binary SN\,II supernovae and/or of the first isolated SN~Ia explosions in the formation of an iron richer second generation;
\item the role of type Ia supernovae to end the epoch of star formation:
\end{itemize}
To further study  this problem, we  show in Fig.~\ref{figure:fig6}  the mass vs. time evolution of super--AGB and AGB models, together with some limiting epochs for important events. We stress again that the timing of different epochs is based on the evolutionary times  of the models we are using throughout this paper, by \cite{ventura2013}, and that the limits may be shifted if other models are adopted, while the general description of the epochs remains qualitatively similar. 

\subsection{The SN\,II epoch boundary (boundaries) }
\label{SNII}

The first (red) boundary in time in Fig.~\ref{figure:fig6}  divides the epoch of single stars SN\,II explosions from the epoch of the quiet AGB evolution. A few further explosions may occur close in time to this limit, if in the most massive super--AGB stars the convective mantle is not fully consumed  before e-captures begin inside the core \citep[e.g.][]{siesspumo2006, poelarendes2008}. 
No cooling flow is possible earlier than this boundary.  

In principle, SN\,II explosions may continue also at a later time, if the critical mass for  explosion (in our example, M$_{\rm  mass}$=8\,\Msun\footnote{We adopt here the definition of \mmass\ as the mass limit for core collapse supernova from \cite{doherty2015}.}) is reached by mass transfer during binary evolution. The existence of this kind of evolutionary path is necessary to explain the presence of young non--recycled pulsars in eccentric orbit with a companion white dwarf \citep{vankerkwijk-kulkarni1999}. In these systems the neutron star formation must have occurred {\it after} the formation of the white dwarf, through mass accretion on the lighter companion \citep{pgzyungelson1999, tauris-sennels2000}. \cite{dantona2005} suggested that binary SN\,II may  influence the formation of multiple populations in GCs. Here we use the simple minded assumption made in this latter work to put a strict time limit to binary SN\,II explosions, by noticing that the more massive (primary) star evolving in the binary  may be assumed, in the best case, to transfer all its envelope, leaving a white dwarf remnant of mass M$_{\rm WD}$.  Until the evolving primary M$_1$\ is massive enough, mass transfer may push the mass accreting component beyond  M$_{\rm mass}$. The end of such delayed SN\,II is set by the time of evolution of the minimum possible mass, which, in the best case, is  M$_1>$(M$_{\rm mass}$+ M$_{\rm WD}$)/2. The limiting time for the possible occurrence of these events in our stellar models is then set by the evolution of the 4.5\msun.  This is a very naif schematization, for a more complete description see, e.g. \cite{tauris-sennels2000}.

An estimate of the impact of binary evolution on late SN\,II explosion is out of scope of the present work. Both the primordial binary fraction for intermediate mass stars of the first generation, their orbital period distribution, and the different possible paths to achieve M$_2$(final)$>$ M$_{\rm mass}$ \citep{tutukovyungelson1993, tauris-sennels2000} are critical in determining the extent of the role played by late SN\,II during the second-generation formation. 
Also the time span of these possible events is strongly dependent on the precise value of \mmass, which not only sets the age of the end of the single SN\,II era, but also determines the {\it minimum} donor mass (and maximum age) which can provide this kind of evolution with mass exchange. 

The actual number and frequency of delayed SN\,II in a cluster strongly depends on many different parameters.
In some cases, they may occur already during the evolution of the highest AGB--superAGB masses, which provide the strongest nuclear processing by p--captures. Thus, the abundance patterns of population E and D in NGC\,2808 require that such explosions did not perturb the cooling flow in this cluster, but they may have been important in clusters which show mild O--Na anticorrelations such as M4 \citep{marino2011m4}.
On the contrary, in NGC\,2419, for which the results of \citet{dicrisci2015} and \citet{cohenkirby2012} suggest that the second generation is entirely or mostly  `extreme' (i.e. made up from undiluted ejecta)  it may be possible that, after this extreme population is formed, an intense delayed SN\,II  epoch begins, followed, without interruptions, by the SN\,Ia epoch. In this case, the  intermediate population born from reaccretion of pristine gas diluting the ejecta is not present at all.

We will show that the most interesting case may be an intermediate one: the observed properties of multiple populations of an entire category of clusters may be explained in a simple way, if the delayed SN\,II explosions indeed have occurred and inhibited star formation for a few tens of million years (\S~\ref{sec:delSNII}).

\subsection{The super--AGB epoch end and the formation of SN\,Ia progenitors }
\label{sec:sn1aform}
The subdividing (blue) line between super--AGB and AGB evolution in Fig.~6 marks the formation of the first C--O white dwarfs. Mass transfer in binaries which contain a mass accreting remnant of super AGBs, an O--Ne core white dwarf, may cause electron capture supernovae \citep[e.g.][]{miyajinomoto1987, gutierrez1996}, but these events should be much less energetic than SN\,Ia \citep{dessart2006}. A very quiet epoch in the cluster life, during which the postulated star formation from super--AGB and AGB ejecta may occur, {\it in the absence of any other perturbing energy sources}, may plausibly exist in clusters.
 
The mechanisms proposed for SN\,Ia explosion all involve one or two C--O white dwarfs. Historically, two models have been discussed, the  single-degenerate \citep[SD, e.g.][]{nomoto1982} and double-degenerate \citep[DD, e.g.][]{webbink1984, ibentutukov1984} model. In the DD model, the SN\,Ia is caused by the merger of two C-O WDs, the combined mass of which equals or exceeds the Chandrasekhar mass. The detonation mechanism of DDs has been debated for decades,
and today rotation is considered the key ingredient in models of this process. Rotation allows mass accumulation up to the explosive central ignition of carbon, which in turn leads to a SN\,Ia \citep{piersanti2003}. The DD case occurs earlier during the galaxies evolution.

Consequently, the  SN\,Ia  epoch can not precede the time at which the first  massive C--O white dwarfs form\footnote{
Recent works have examined the possibility that Carbon burning rates are very different (either larger or smaller) from the recommended values \citep[e.g.][]{chen2014}. This assumption leads to the formation of O--Ne--C cores which would also explode as SN\,Ia if their mass increases up to the Chandrasekhar mass by mass transfer. This  is still only a working hypothesis, so we assume it does not alter our main conclusion.}, plus a further time delay necessary to achieve the mass transfer by which this white dwarf attains the explosion conditions \citep[e.g.][]{madau1998}.   
Cluster-to-cluster variations in the onset of the SN\,Ia epoch will lead to differences in the star formation modalities of the latest multiple populations. In particular, it is even possible that the SG formation stops  {\it before} the CNO and s--process enhanced ejecta begin to be dominant. 

\subsection{The SN~Ia epoch and the (final) end of SG star formation }
\label{type1a}
The observational constraints on SN~Ia frequency, and its probable starting epoch \citep{mannucci2005, totani2008} are based on observations of different types of galaxies, so it is necessary to understand how these findings may be transferred to the study of globular clusters. \\
Recent observations have shown that the peak of frequency of events is, at early times, just somewhat above 10$^8$yr, in galaxies still subject to star formation \citep{mannucci2005, totani2008}. \cite{totani2008} study shows the beginning of the SN\,Ia data at a ``delay time" interval 100--250\,Myr.\footnote{The delay time here is is ``the delay time from star formation", so it is the total age, if we are dealing with a burst of star formation. In this nomenclature, the delay time in Fig.\,6 is 10$^8$\,yr, while the true delay time between the first CO white dwarf formation and the SN~Ia epoch is (100--65)=35\,Myr. } In this context, the DD model is able to provide large rates at early times \citep{mennekens2010}, 
and this might be particularly true in Globular Clusters, where we can expect to have many close binaries with two CO white dwarfs components, which can merge after a short delay time, thanks to the spiral-in caused by the emission of gravitational wave radiation. 

So the timescale for the onset of the SN~Ia epoch in early globular clusters is in the same age range of the end of the binary delayed SN~II epoch, and in the middle of the CNO-- and s--enhanced AGB ejecta. Possible ---and probable--- variations in the role of these ingredients are likely to affect cluster-to-cluster differences in the properties of multiple populations.

\subsection{The epoch of C+N+O and s--process enriched AGB ejecta}
Fig.~6 shows that the timing of  production of CNO-- and s--process-- rich AGB ejecta, the delayed SN~II epoch and the beginning of the SN\,Ia epoch are partially overlapping, so that their roles can not be examined separately.  On the other hand, clusters in which CNO and s--process enhancement is present do exist.

In the early years of observations of chemical anomalies in GCs, it appeared that  the total C+N+O abundance was constant in the individual clusters examined, and that the C, N and O variations were to be ascribed to the action of the CNO cycle, and in particular of the ON branch, which accounted for the oxygen reduction in the second generation stars. Examples of such clusters are M92: \cite{pilac1988}; NGC~288 and NGC~362: \cite{dickens1991};  M~3 and M~13: \cite{smith1996}; M4: \cite{ivans1999}; NGC~6752 \citep{yong2015}.  

More recently, the situation became more complex. First, a simple explanation of the splitting found in the subgiant branch of the cluster NGC~1851 \citep{milone2008} was more easily justified by assuming that its SG (populating the dimmer branch) had a larger CNO than the bright branch, but a similar age \citep{cassisi2008,  ventura18512009}. Several other clusters have been shown to hold a double sub giant branch \citep{piotto2012}. The spectroscopic enrichment in total CNO is established only in NGC~1851 \citep{yong2009, yong2015},   and M22 \citep{marino2012m22}, see also \cite{ lim2015b}\footnote{\ocen\ too has an increase in s--process and CNO abundances \citep{marino2012cno}, but its evolution may have been more complex anyway.}.  In clusters showing a split sub giant branch, this  feature is accompanied by an increase in s-process elements abundance, which is in line with the 3DU interpretation \citep{straniero2014}. 

Further investigations are necessary, to explore the processes which may explain why some clusters show two main bursts of SF, separated by a few tens of million years to allow the CNO--, s--enhanced ejecta production.

\section{Clusters with iron spread or bimodality}
Recent spectroscopic observations have found a few GCs characterized by an internal [Fe/H] abundance dispersion.
As suggested by \cite{dacosta2015}, these systems might be the former nuclear star clusters of now disrupted
dwarf galaxies \citep{willmanstrader2012, marino2015}. An iron difference among cluster stars implies that the cluster was
able to retain at least some of the supernova ejecta. This may occur either in dwarf galaxies, where possibly dark matter is initially
present, or in particularly massive clusters.  The spread in iron observed ranges from tiny differences, as in NGC\,1851 to the very large differences observed in  $\omega$\,Cen. 

In all these clusters, the typical signature of p--capture elements variations (the O--Na anticorrelation) are accompanied by metal enrichment, CNO  and s--process enrichment together. The category of clusters named ``s--Fe--anomalous" described by \cite{marino2015} includes mainly clusters with split sub giant branch, namely NGC~1851 \citep{carretta2011,gratton2012sgbs1851}, NGC~5286 \citep{marino2015}, M~22 \citep{marino2009} and M2 \citep{yong2014}, plus \ocen\ \citep{johnson2009, marino2011wcen}. 

\subsection{Pollution by delayed SN~II: the s--Fe--anomalous clusters}
\label{sec:delSNII}
Here we propose how a natural extension of the standard GC model for the formation of multiple populations can deal with the s-Fe-anomalous clusters. In these clusters, the O--Na anticorrelation is already present in the fraction of stars having the smaller, and homogeneous, iron content. This anticorrelation is very unlikely to occur, if the iron contamination is due to SN\,II of the single SN\,II epoch. So we suggest that the first phases of evolution of these clusters are similar to what happens in GCs which are fully chemically homogeneous in heavy elements, but, at later times, we must account for contamination by further supernova explosions,  those from delayed SN\,II in binaries. Let us assume: 
\begin{enumerate}
\item delayed SN\,II do not explode before the  first SG, homogeneous in iron, forms, providing a typical own O--Na anti correlation;  
\item afterwards, delayed SN\,II begin exploding with some regularity in a cluster, destroy the cooling flow, which would have included both AGB ejecta and pristine gas;
\item these events (which are much less frequent than the single SN\,II were) are not able to inject into the gas enough power to definitely push it out of the cluster vicinity. 
\end{enumerate}
When, at last,  the delayed events become rare, possibly several tens of Myr later, the pristine plus AGB gas will re-accrete and induce a new SG formation burst. In these hypotheses, {\it the pristine gas will now be contaminated by the delayed SN\,II ejecta, and by the AGB ejecta which were lost during this time span}. Further, the contaminating AGBs will be the masses in which the 3DU has been very effective, so this population will have the characteristics of the ``s--Fe--anomalous" clusters: larger iron and s--process abundances, and associated C+N+O enhancement. 

An appealing feature of this scenario is that it explains the features of the anomalous clusters, without the need for additional hypotheses, such as the merging of two  different clusters. Dynamical models are anyway required to test this suggestion, and are under way (D'Ercole et al. 2015, submitted to MNRAS). 

The time gap of a few 10$^7$yr  between the formation of the `first', standard SG, and the `second' one, s-Fe and CNO enriched, also justifies another important characteristics of some clusters: the presence of separate subgiant branches. 
While this time break is negligible in terms of location of isochrones with identical chemical composition, this short time is sufficient to shift the AGB ejecta composition to the CNO enriched stage, which, also with the help of the small iron increase, will result in distinct subgiant branches \citep[for the case of NGC~1851, see][]{cassisi2008, ventura18512009}.
Finally notice that the formation epoch of this s-Fe-CNO enriched SG can not be very extended, as it occurs close to the beginning of the SN\,Ia era, which will definitely end star formation. 
\\
\begin{table*}
	\centering
	\caption{Mass polluted by one SN~Ia (or by 10 binary delayed SN~II) versus Fe--anomalous masses in GCs }
	\label{tab:table2}
	\begin{tabular}{ccccccccccc} 
	\hline \hline
Cluster  & M/M$_\odot ^1$   &  [Fe/H]$_{\rm in}$  &   [Fe/H]$_{\rm fin}$  & \%   &  M$_{\rm poll.expected}$  & M$_{\rm poll.obs}$ & $\Delta$M/M $^2$ & $\delta$[Fe/H]  & Nomenclature & Ref$^{3}$ \\ 
\hline \\
NGC~2808  & 8.5e5 &  --1.1  &  -1.0  &  0.056  & 4.8$\times 10^4$   &   4.8$\times 10^4$  & 0      & 0.1 &  pop.A Mil15a &this paper \\
 M2  & 6.9e5 &  --1.7  &  -1.5 &  0.03  &  8.7$\times 10^4$ &  2.1$\times 10^4$  & +0.76   & 0.2 &  pop.BI+BII Mil15b &  Mil15\\
 M2  & 6.9e5 &  --1.7  &  -1.0 &  0.01  & 1.2$\times 10^4$ &   6.9$\times 10^3$  & +0.42 & 0.7 &  pop. C Mil15b&  Mil15b, Y14 \\

NGC~1851  & 3.1e5 &  --1.2  &  -1.15 &  0.45  & 1.4$\times 10^5$  &  2.5$\times 10^5$  & --0.1 & 0.05 &   faint SGB &  Mil08 \\
NGC~5286  & 4.5e5 &  --1.8  &  -1.6 &  0.15  & 6.7$\times 10^4$   & 7.2$\times 10^4$  &+0.33 & 0.2 &   faint SGB &  Mar15 \\
 M22   & 3.6e5 &  --1.82  &  -1.67 &  0.35  &  1.3$\times 10^5$  & 1.9$\times 10^5$  & +0.28 & 0.15 &   faint SGB &  Mil12 \\
\hline \hline
\multicolumn{7}{l}{$^1$ cluster masses from \cite{mclvdm2005massgc}}\\
\multicolumn{7}{l}{$^2$ $\Delta$M/M=(M$_{\rm poll. expected}$--M$_{\rm observed}$)/M$_{\rm poll. expected}$} \\
\multicolumn{11}{l}{$^{3}$ Mil15a: Milone et al. 2015a; Mil15b: Milone et al. 2015b; Mil08: Milone et al. 2008; Y14: Yong et al. 2014} \\
\end{tabular}
\end{table*}

\subsection{Are there signatures of isolated episodes of delayed SN\,II explosions?}
We remark here that there might be clusters in which one isolated episode, or a few separated episodes of delayed SN~II explosions occur. In this case, the explosion is able to halt only partially the cooling flow, and some of the gas flowing into the cluster core will be contaminated by the ejecta of this or these supernovae. This is an additional possibility to be taken into account when dealing with the formation of populations with abundance anomalies. Each single type II supernova can not alter the iron content of the gas in a detectable way, but it can certainly increase the oxygen content, and also the Mg and Si content. The s--process and the C+N+O may result increased or not for events which occur later or earlier during the time span allowed for the delayed SN~II events. \\
It would be important to carry out observational studies aimed at searching the fingerprints of this scenario; signatures of O and Mg increase may be revealed by the combination of the UV and optical HST passbands of the UV Legacy Survey of Galactic GCs \citep{piotto2015}; studies of the iron abundance will require very accurate datasets and analysis, such as those of \cite{yong2013} for the cluster NGC~6752, where a spread of 0.03~dex in [Fe/H] has been detected, while no spread in CNO seems to be present in this cluster \citep{yong2015}.

\subsection{Signatures from the SN~Ia first random explosions}
\label{sec:sn1a}
The SN~Ia, unlike these delayed SNII, play a very different (sometimes double) role. In fact:

1) first they might pollute the reaccreting gas, producing a more metal rich last population as we suggest for NGC~2808;  

2) when the frequency of explosions becomes stationary (at an age of $\gtrsim 10^8$yr), they are likely to completely halt the star formation. \\
Before reaching a frequency of events large enough to halt the cooling flow and the second generation formation, a few  initial SN\,Ia explosions  might occur well separated in time each other. \cite{dercole2008} have shown that one SN\,Ia event, isolated in time, is not able to halt the cooling flow. A few events may be needed, especially if the cluster is suffering strong gas reaccretion. While this problem is to be further studied in more detail, we think it is likely to lead to a variety of different outcomes.  NGC\,2808 remains a valid example possibility: we expect that pollution by iron from SN\,Ia in the pristine reaccreting gas (mixed, or not, with AGB ejecta) will result in the formation of stars which will be more metal rich. 

\begin{figure}
\centering{
\vskip -30pt
\includegraphics[width=9cm]{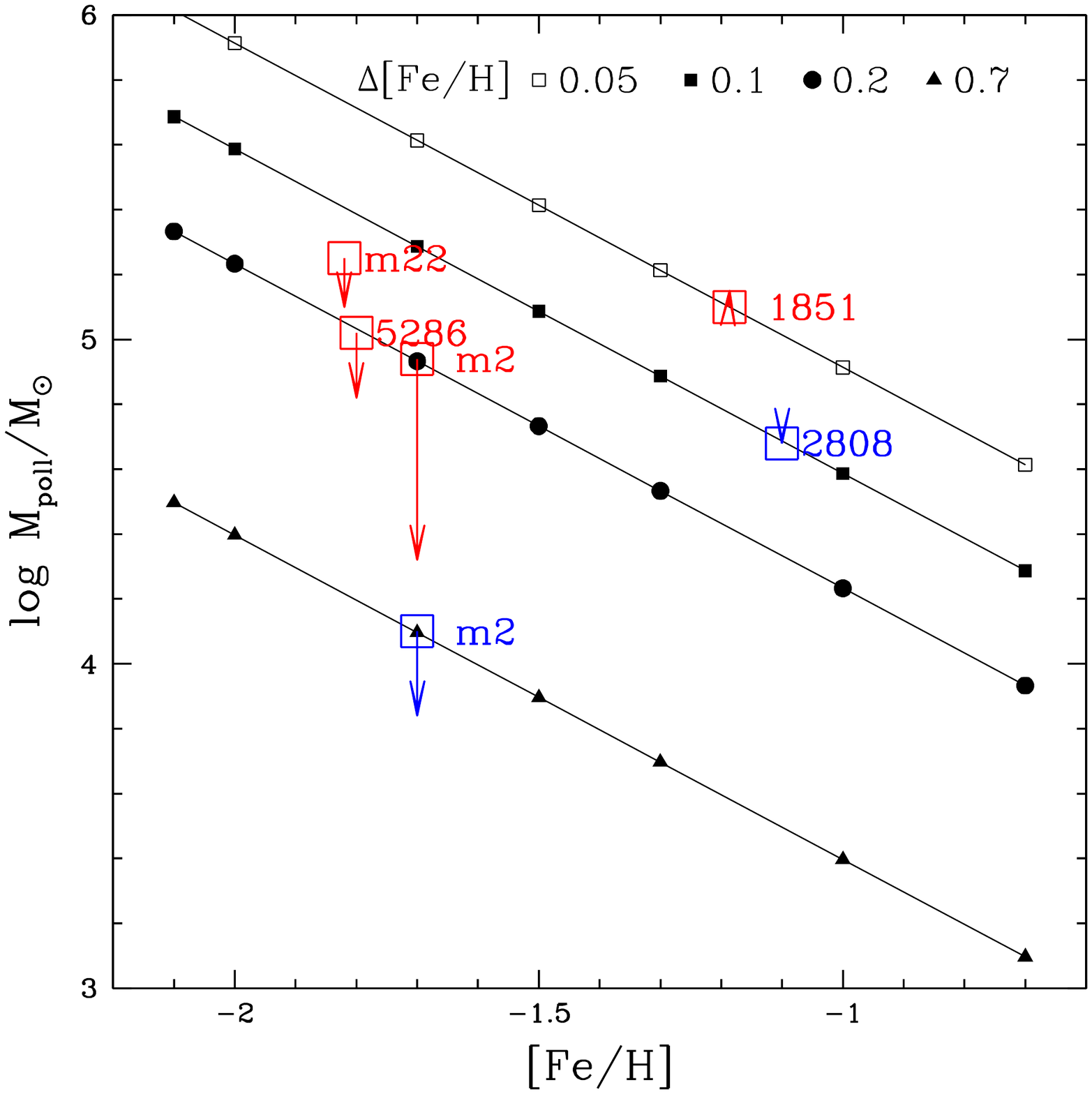}
\vskip -50pt
\caption{The lines represent the mass of gas which can be polluted by 1 SN\,Ia exploding in a cluster of metallicity [Fe/H], to achieve a  metallicity larger by $\Delta$[Fe/H], for the values of $\Delta$[Fe/H] labelled at the top. We assume that the SN~Ia provides 0.8\msun\ of iron. We show where the populations with increased iron present in some clusters are located in this diagram, based on each cluster [Fe/H] and on the average $\Delta$[Fe/H]  of the population under study. For the same clusters, we also estimate the mass of this population with enhanced iron content from the percentage of cluster stars it contains and the cluster mass. The tip of the arrows on each square indicate where this actual mass is with respect to the evaluated mass (see Table 3 for details). The estimated masses coincide or are larger than the actual masses, indicating that indeed one SN~Ia may have polluted the gas from which the population formed. However, we suggest that the only cases in which pollution comes from the first SN~Ia explosion are those represented by the blue boxes: NGC\,2808 pop. A in Milone et al. 2015a and M2 pop. C in Milone et al. 2015b. }
\label{fig7}
}
\end{figure}

\section{A zero--order estimate of pollution by one SN~Ia or by delayed SN~II}
For NGC\,2808 we have estimated the amount of gas which can be polluted by a single SN\,Ia, assuming that the average $\delta$[Fe/H] is 0.1\,dex larger than the iron content of the rest of cluster stars. This leads to a prediction of  48000\msun\ of polluted gas, in very good agreement with the 5.6\% of the total mass of the cluster, estimated for the `A' population by \cite{milone2015} \citep[about 47700\msun\ if we assume 8.5$\times 10^5$\msun\ for the entire cluster,][]{mclvdm2005massgc}. \\
\begin{table*}
	\centering
	\caption{Timeline for the formation of multiple populations in different clusters}
	\label{tab:table3}
	\begin{tabular}{ccccccccc} 
	\hline
time(Myr)       & 0--3      & 5--40                   &  40--60     &  60--90 & 90--100 & 100--120--? & >200  \\
 M/M$_\odot$ &&   $\rightarrow$ 8  &   8 -- 6.5      &  6.5 -- 5   &  $\sim$5   &  5--  4.   &  $<$3   \\
		 \hline
 population        &  FG   &  no SF    &   Extreme  &  Intermediate   & \multicolumn{2}{c}{late--Interm., CNO, s and Fe enriched}   &  \\
		\hline
 events            &  FG       &    SN\,II  &  SAGB & massive AGB  &          
CNO $\uparrow$   &  CNO $\uparrow$ $\uparrow$    $\rightarrow$ &  C-star type ejecta  \\  

                     &&&  & s-process $\uparrow$   &  s-process $\uparrow$ $\uparrow$   $\rightarrow$ & Yes/No \\
                      &       &   &   \multicolumn{4}{c}{$\leftarrow\leftarrow\leftarrow\leftarrow\leftarrow\leftarrow$ ~~~~~~~~~~~ SN\,II ~~~~~~~ $\rightarrow\rightarrow\rightarrow\rightarrow\rightarrow\rightarrow\rightarrow\rightarrow$}  &  &      \\  
                     &&&    \multicolumn{4}{c}{delayed}  &  &    \\

                     &   &    & & & \multicolumn{3}{c}{$\rightarrow\rightarrow$ ~ SN\,Ia  $\rightarrow\rightarrow$} &  \\   
                     &  & & \multicolumn{4}{c}{.............................. episodes of re-accretion ..............................} \\   
                     \\
		 \hline                      
 		\hline
\multicolumn{9}{c} {classic clusters}  \\
\hline    
NGC~2808   & FG   &   no SF   &  pure ejecta SF   & diluted gas SF  & strong diluted SF    &  \multicolumn{2}{l}{ first SNIa SF, SNIa epoch} \\ \\

NGC~2419   & FG   &   no SF   &  pure ejecta SF   & \multicolumn{2}{c}{$\leftarrow\leftarrow\leftarrow\leftarrow$  delayed SN\,II epoch  $\rightarrow\rightarrow\rightarrow\rightarrow$}   &  \multicolumn{2}{l}{ SNIa epoch} \\ \\

    M4            & FG   &   no SF   &  \multicolumn{2}{c}{$\leftarrow\leftarrow$ ~~~~~~~~diluted~gas~SF~~~~~ $\rightarrow\rightarrow$}   &  stop (SNIa?)    \\ \\
\hline \hline
\multicolumn{9}{c} {double SGB clusters}  \\
\hline   
M22,          & FG   &   no SF   &  first dilution  &   \multicolumn{1}{c}{delayed SN\,II epoch }  &  fast recollapse, &  SNIa epoch \\
NGC 1851, &  &    &  SF Fe initial  &  no SF &   burst SF, Fe $ \uparrow$ &   \\ 
NGC 6656, &  &    &  &   &  & \\ 
NGC 5286  &  &    &  &   &  & \\ \\
\hline \hline
\multicolumn{9}{c} {triple SGB clusters}  \\
\hline   
M2          & FG   &   no SF   &  first dilution  &  delayed SN\,II epoch  &  fast recollapse, &   \multicolumn{2}{l}{ first SNIa SF, SNIa epoch} \\
& &    &  SF Fe initial  &  no SF   &  burst SF, Fe $ \uparrow$ & SF Fe $ \uparrow\uparrow$&\\
\hline\hline

\end{tabular}
\end{table*}

On the same grounds we may try to understand whether the other populations with larger iron content found in several clusters can be explained by pollution by a single SN~Ia, assumed to eject 0.8\msun\ of iron. Notice that 
the iron mass ejected by a low mass SN~II is about 1/10 of the iron mass of SN~Ia \citep[0.07\msun, according to ][]{nomoto2013ARAA}, so the schematic reasoning which follows can be applied to pollution either by one SN~Ia, or by ten SN~II. Anyway, in order to justify that the iron increase in the whole SG considered has a small spread, we must also think that there has been a global mixing of all the SN~II ejecta with the pristine gas.
\\
We show in Fig.~\ref{fig7} and Table~\ref{tab:table2} the result obtained under these simplified assumptions. We plot the lines of  expected polluted mass as a function of the [Fe/H] of the standard cluster population, and of the $\delta$[Fe/H] of the anomalous population. We compare the predictions with the `observed' mass, knowing the fraction of involved stars and the total mass of each cluster (Table \ref{tab:table2}). The two numbers are in reasonable agreement with the observations. In one case (the populations BI+BII of M2 in Milone et al. 2015) the expected mass is far larger than observed. In the other cases it is comparable or modestly larger. 

Should we then conclude that all clusters showing the presence of an iron enhanced population have experienced contamination by this first SN~Ia ejecta? Actually, we suggest that this idea is reasonable for pop. C in M2 and for pop. A in NGC~2808, for which these groups represent a small fraction of the total cluster mass, and do not show an internal O--Na anticorrelation. 
In all the other cases examined in Table \ref{tab:table2}, the presence of a double SGB suggests that we are dealing with a burst of star formation  after it has been inhibited for a long time, and we find more natural to attribute the iron increase to delayed binary SN\,II, as described in \S~\ref{sec:delSNII}. So, for instance, we estimate that the CNO-- s--rich population in NGC~1851 requires contamination by $\sim$10 delayed SN~II, rather than by one SN~Ia. We point out that, obviously, the number of delayed SN\,II may have been larger than this because not all the matter polluted by these ejecta will be converted in stars during this final star formation event (in particular because this last star formation event may be interrupted by the beginning of the SN\,Ia epoch.)

A small calcium increase in the s-Fe-rich population, found in M~22 \citep{dacosta2009, marino2009} ---see also \cite{lee2009nature}---
in NGC~1851 \citep{han2009} ---see also \cite{lim2015b, lim2015calcium}--- and possibly in the population BI+BII of M2 
\citep{yong2014} is also a signature of SN\,II contamination, although, in other contexts, a calcium variation may be a sign of SN\,Ia contamination.

\section{Conclusions}
We have shown that the AGB scenario provides a timeline along which the large variety of multiple populations in GCs may find its place, as we have described by reconstructing the SFH of the five or more populations born in NGC~2808.  The limitation of this analysis is mostly met when we attempt a quantitative comparison with the abundances of magnesium, silicon and aluminum, but all  the qualitative trends and the general reconstruction of  the events of star formation are adequately met. No other model so far proposed as a basis for the chemical variety of multiple populations can meet such a large number of chemical constraints.

For NGC~2808 our analysis extends the model based on the AGB scenario which nicely accounted for the three populations previously known to be present in the cluster. The extended model is able to naturally account for the presence of two further populations recently identified by \cite{milone2015}, the first one constituting $\sim$25\% of the cluster mass, somewhat enriched in nitrogen, but scarcely enriched in helium and sodium. We show that this group of stars requires that star formation has occurred in gas formed by AGB ejecta in which C+N+O is larger than the initial value, due to the effects of the 3DU, very diluted with  pristine gas. We also attempt to explain the small population `A' isolated in \cite{milone2015}, by assuming that a last episode of star formation involves gas polluted by the iron--rich ejecta of the first SN~Ia explosion.

In order to extend the star formation history to other clusters, we have examined all the different epochs which may occur in clusters in the first 100--120\,Myr of life, in particular we have re--examined the possible role of delayed SN~II explosions occurring in binaries in which mass exchange has risen the secondary component of the system above M$_{\rm mass}$. We show that a limited number of explosions lasting for several tens of Myr  may stop the star formation until these supernovae explode, and lead finally to an intense burst of star formation which may explain the characteristics of s-Fe-anomalous clusters. 

In Table~\ref{tab:table3} we summarize schematically our proposal for the different events taking place in different prototype clusters.

While further investigations aimed at exploring the occurrence of the events leading to cluster-to-cluster differences are necessary, the scenario we present, although based on imperfect stellar models, can explain the large variety of multiple populations features in clusters. \\
In Table~\ref{tab:table3} we do not include \ocen, because, in our scheme, the initial phases of its evolution can not account for the fast s--process increase in the low metallicity range \citep{johnson2009, marino2011wcen, dantona2011wcen}, but we point out that the formation of its metal richer populations may have followed paths similar to those proposed here.

Much work remains to be done, and the scenario has enough details that it can be falsified by further analysis.
The AGB nucleosynthesis is still very uncertain and the details of the proposed scheme(s) are not fully settled, but a whole and variegated set of results shows a global consistency with models whose temporal evolution is reasonably in line with what we know about stellar and dynamical evolution in clusters. We claim that, at this stage, the formation of multiple stellar generations in clusters, from matter including massive AGB ejecta, is the only viable option.

\section*{Acknowledgements}
We thank the referee for a useful report which helped to improve the clarity of the work.
We thank Eugenio Carretta for useful comments and Lev Yungelson, Nicki Mennekens and Amedeo Tornambe\`\  for discussions on supernova rates. We are indebted to Gianluca Imbriani for enlightenment on the problems of the sodium plus proton reaction channels.
F.D'A.  and A.D'E acknowledge support from PRIN INAF 2014 (principal investigator S. Cassisi). 
A.P.M. acknowledges support by the Australian Research Council through Discovery Early Career Researcher Award DE150101816. A.F.M. acknowledges support by the Australian Research Council through Discovery Early Career Researcher Award DE160100851. E.V. acknowledges support from grant NASA-NNX13AF45G.  



\bsp	


\begin{thebibliography}{126}
\expandafter\ifx\csname natexlab\endcsname\relax\def\natexlab#1{#1}\fi

\bibitem[{{Bastian} {et~al.}(2015){Bastian}, {Cabrera-Ziri}, \&
  {Salaris}}]{bastian2015}
{Bastian} N., {Cabrera-Ziri} I., {Salaris} M., 2015, \mnras, 449, 3333

\bibitem[{{Bastian} {et~al.}(2013){Bastian}, {Lamers}, {de Mink}, {Longmore},
  {Goodwin}, \& {Gieles}}]{bastian2013}
{Bastian} N., {Lamers} H.~J.~G.~L.~M., {de Mink} S.~E., {Longmore} S.~N.,
  {Goodwin} S.~P., {Gieles} M., 2013, \mnras, 436, 2398

\bibitem[{{Becker} \& {Iben}(1979)}]{beckeriben1979}
{Becker} S.~A., {Iben} Jr. I., 1979, \apj, 232, 831

\bibitem[{{Behr}(2003)}]{behr2003}
{Behr} B.~B., 2003, \apjs, 149, 67

\bibitem[{{Bekki}(2011)}]{bekki2011}
{Bekki} K., 2011, \mnras, 412, 2241

\bibitem[{{Bloecker}(1995)}]{blocker1995}
{Bloecker} T., 1995, \aap, 297, 727

\bibitem[{{Boothroyd} {et~al.}(1993){Boothroyd}, {Sackmann}, \&
  {Ahern}}]{bs1993}
{Boothroyd} A.~I., {Sackmann} I.-J., {Ahern} S.~C., 1993, \apj, 416, 762

\bibitem[{{Bressan} {et~al.}(1993){Bressan}, {Fagotto}, {Bertelli}, \&
  {Chiosi}}]{bressan1993}
{Bressan} A., {Fagotto} F., {Bertelli} G., {Chiosi} C., 1993, \aaps, 100, 647

\bibitem[{{Busso} {et~al.}(1999){Busso}, {Gallino}, \&
  {Wasserburg}}]{busso1999ARAA}
{Busso} M., {Gallino} R., {Wasserburg} G.~J., 1999, \araa, 37, 239

\bibitem[{{Canuto} {et~al.}(1996){Canuto}, {Goldman}, \&
  {Mazzitelli}}]{cgm1996}
{Canuto} V.~M., {Goldman} I., {Mazzitelli} I., 1996, \apj, 473, 550

\bibitem[{{Canuto} \& {Mazzitelli}(1991)}]{cm1991}
{Canuto} V.~M., {Mazzitelli} I., 1991, \apj, 370, 295

\bibitem[{{Carretta}(2014)}]{carretta2014}
{Carretta} E., 2014, \apjl, 795, L28

\bibitem[{{Carretta}(2015)}]{carretta2015}
---, 2015, \apj, 810, 148

\bibitem[{{Carretta} {et~al.}(2009{\natexlab{a}}){Carretta}, {Bragaglia},
  {Gratton}, {D'Orazi}, \& {Lucatello}}]{carretta2009c}
{Carretta} E., {Bragaglia} A., {Gratton} R., {D'Orazi} V., {Lucatello} S.,
  2009{\natexlab{a}}, \aap, 508, 695

\bibitem[{{Carretta} {et~al.}(2006){Carretta}, {Bragaglia}, {Gratton}, {Leone},
  {Recio-Blanco}, \& {Lucatello}}]{carretta2006}
{Carretta} E., {Bragaglia} A., {Gratton} R.~G., {Leone} F., {Recio-Blanco} A.,
  {Lucatello} S., 2006, \aap, 450, 523

\bibitem[{{Carretta} {et~al.}(2009{\natexlab{b}}){Carretta}, {Bragaglia},
  {Gratton}, {Lucatello}, {Catanzaro}, {Leone}, {Bellazzini}, {Claudi},
  {D'Orazi}, {Momany}, {Ortolani}, {Pancino}, {Piotto}, {Recio-Blanco}, \&
  {Sabbi}}]{carretta2009a}
{Carretta} E., {Bragaglia} A., {Gratton} R.~G., {Lucatello} S., {Catanzaro} G.,
  {Leone} F., {Bellazzini} M., {Claudi} R., {D'Orazi} V., {Momany} Y.,
  {Ortolani} S., {Pancino} E., {Piotto} G., {Recio-Blanco} A., {Sabbi} E.,
  2009{\natexlab{b}}, \aap, 505, 117

\bibitem[{{Carretta} {et~al.}(2012){Carretta}, {Bragaglia}, {Gratton},
  {Lucatello}, \& {D'Orazi}}]{carretta2012}
{Carretta} E., {Bragaglia} A., {Gratton} R.~G., {Lucatello} S., {D'Orazi} V.,
  2012, \apjl, 750, L14

\bibitem[{{Carretta} {et~al.}(2013){Carretta}, {Gratton}, {Bragaglia},
  {D'Orazi}, {Lucatello}, {Sollima}, \& {Sneden}}]{carretta2013}
{Carretta} E., {Gratton} R.~G., {Bragaglia} A., {D'Orazi} V., {Lucatello} S.,
  {Sollima} A., {Sneden} C., 2013, \apj, 769, 40

\bibitem[{{Carretta} {et~al.}(2011){Carretta}, {Lucatello}, {Gratton},
  {Bragaglia}, \& {D'Orazi}}]{carretta2011}
{Carretta} E., {Lucatello} S., {Gratton} R.~G., {Bragaglia} A., {D'Orazi} V.,
  2011, \aap, 533, A69

\bibitem[{{Cassisi} {et~al.}(2008){Cassisi}, {Salaris}, {Pietrinferni},
  {Piotto}, {Milone}, {Bedin}, \& {Anderson}}]{cassisi2008}
{Cassisi} S., {Salaris} M., {Pietrinferni} A., {Piotto} G., {Milone} A.~P.,
  {Bedin} L.~R., {Anderson} J., 2008, \apjl, 672, L115

\bibitem[{{Chantereau} {et~al.}(2015){Chantereau}, {Charbonnel}, \&
  {Decressin}}]{chantereau2015}
{Chantereau} W., {Charbonnel} C., {Decressin} T., 2015, \aap, 578, A117

\bibitem[{{Chen} {et~al.}(2014){Chen}, {Herwig}, {Denissenkov}, \&
  {Paxton}}]{chen2014}
{Chen} M.~C., {Herwig} F., {Denissenkov} P.~A., {Paxton} B., 2014, \mnras, 440,
  1274

\bibitem[{{Clement} \& {Hazen}(1989)}]{clement1989rr2808}
{Clement} C.~M., {Hazen} M.~L., 1989, \aj, 97, 414

\bibitem[{{Cohen} \& {Kirby}(2012)}]{cohenkirby2012}
{Cohen} J.~G., {Kirby} E.~N., 2012, \apj, 760, 86

\bibitem[{{Conroy}(2012)}]{conroy2012}
{Conroy} C., 2012, \apj, 758, 21

\bibitem[{{Cordero} {et~al.}(2015){Cordero}, {Pilachowski}, {Johnson}, \&
  {Vesperini}}]{cordero2015}
{Cordero} M.~J., {Pilachowski} C.~A., {Johnson} C.~I., {Vesperini} E., 2015,
  \apj, 800, 3

\bibitem[{{Costantini} {et~al.}(2009){Costantini}, {Formicola}, {Imbriani},
  {Junker}, {Rolfs}, \& {Strieder}}]{luna2009}
{Costantini} H., {Formicola} A., {Imbriani} G., {Junker} M., {Rolfs} C.,
  {Strieder} F., 2009, Reports on Progress in Physics, 72, 086301

\bibitem[{{Da Costa}(2015)}]{dacosta2015}
{Da Costa} G.~S., 2015, ArXiv e-prints (arXiv:1510.00873)

\bibitem[{{Da Costa} {et~al.}(2009){Da Costa}, {Held}, {Saviane}, \&
  {Gullieuszik}}]{dacosta2009}
{Da Costa} G.~S., {Held} E.~V., {Saviane} I., {Gullieuszik} M., 2009, \apj,
  705, 1481

\bibitem[{{Dalessandro} {et~al.}(2011){Dalessandro}, {Salaris}, {Ferraro},
  {Cassisi}, {Lanzoni}, {Rood}, {Fusi Pecci}, \& {Sabbi}}]{dalessandro2011}
{Dalessandro} E., {Salaris} M., {Ferraro} F.~R., {Cassisi} S., {Lanzoni} B.,
  {Rood} R.~T., {Fusi Pecci} F., {Sabbi} E., 2011, \mnras, 410, 694

\bibitem[{{D'Antona} {et~al.}(2005){D'Antona}, {Bellazzini}, {Caloi}, {Pecci},
  {Galleti}, \& {Rood}}]{dantona2005}
{D'Antona} F., {Bellazzini} M., {Caloi} V., {Pecci} F.~F., {Galleti} S., {Rood}
  R.~T., 2005, \apj, 631, 868

\bibitem[{{D'Antona} \& {Caloi}(2004)}]{dc2004}
{D'Antona} F., {Caloi} V., 2004, \apj, 611, 871

\bibitem[{{D'Antona} \& {Caloi}(2008)}]{dc2008}
---, 2008, \mnras, 390, 693

\bibitem[{{D'Antona} {et~al.}(2011){D'Antona}, {D'Ercole}, {Marino}, {Milone},
  {Ventura}, \& {Vesperini}}]{dantona2011wcen}
{D'Antona} F., {D'Ercole} A., {Marino} A.~F., {Milone} A.~P., {Ventura} P.,
  {Vesperini} E., 2011, \apj, 736, 5

\bibitem[{{D'Antona} \& {Ventura}(2007)}]{dantonaventura2007}
{D'Antona} F., {Ventura} P., 2007, \mnras, 379, 1431

\bibitem[{{de Mink} {et~al.}(2009){de Mink}, {Pols}, {Langer}, \&
  {Izzard}}]{demink2009}
{de Mink} S.~E., {Pols} O.~R., {Langer} N., {Izzard} R.~G., 2009, \aap, 507, L1

\bibitem[{{Decressin} {et~al.}(2007){Decressin}, {Meynet}, {Charbonnel},
  {Prantzos}, \& {Ekstr{\"o}m}}]{decressin2007}
{Decressin} T., {Meynet} G., {Charbonnel} C., {Prantzos} N., {Ekstr{\"o}m} S.,
  2007, \aap, 464, 1029

\bibitem[{{Denissenkov} \& {Hartwick}(2014)}]{denissenkov2014}
{Denissenkov} P.~A., {Hartwick} F.~D.~A., 2014, \mnras, 437, L21

\bibitem[{{D'Ercole} {et~al.}(2012){D'Ercole}, {D'Antona}, {Carini},
  {Vesperini}, \& {Ventura}}]{dercole2012}
{D'Ercole} A., {D'Antona} F., {Carini} R., {Vesperini} E., {Ventura} P., 2012,
  \mnras, 423, 1521

\bibitem[{{D'Ercole} {et~al.}(2010){D'Ercole}, {D'Antona}, {Ventura},
  {Vesperini}, \& {McMillan}}]{dercole2010}
{D'Ercole} A., {D'Antona} F., {Ventura} P., {Vesperini} E., {McMillan}
  S.~L.~W., 2010, \mnras, 407, 854

\bibitem[{{D'Ercole} {et~al.}(2008){D'Ercole}, {Vesperini}, {D'Antona},
  {McMillan}, \& {Recchi}}]{dercole2008}
{D'Ercole} A., {Vesperini} E., {D'Antona} F., {McMillan} S.~L.~W., {Recchi} S.,
  2008, \mnras, 391, 825

\bibitem[{{Dessart} {et~al.}(2006){Dessart}, {Burrows}, {Ott}, {Livne}, {Yoon},
  \& {Langer}}]{dessart2006}
{Dessart} L., {Burrows} A., {Ott} C.~D., {Livne} E., {Yoon} S.-C., {Langer} N.,
  2006, \apj, 644, 1063

\bibitem[{{Di Criscienzo} {et~al.}(2011){Di Criscienzo}, {D'Antona}, {Milone},
  {Ventura}, {Caloi}, {Carini}, {D'Ercole}, {Vesperini}, \&
  {Piotto}}]{dicrisci2011}
{Di Criscienzo} M., {D'Antona} F., {Milone} A.~P., {Ventura} P., {Caloi} V.,
  {Carini} R., {D'Ercole} A., {Vesperini} E., {Piotto} G., 2011, \mnras, 414,
  3381

\bibitem[{{Di Criscienzo} {et~al.}(2015){Di Criscienzo}, {Tailo}, {Milone},
  {D'Antona}, {Ventura}, {Dotter}, \& {Brocato}}]{dicrisci2015}
{Di Criscienzo} M., {Tailo} M., {Milone} A.~P., {D'Antona} F., {Ventura} P.,
  {Dotter} A., {Brocato} E., 2015, \mnras, 446, 1469

\bibitem[{{Dickens} {et~al.}(1991){Dickens}, {Croke}, {Cannon}, \&
  {Bell}}]{dickens1991}
{Dickens} R.~J., {Croke} B.~F.~W., {Cannon} R.~D., {Bell} R.~A., 1991, \nat,
  351, 212

\bibitem[{{Doherty} {et~al.}(2014){Doherty}, {Gil-Pons}, {Lau}, {Lattanzio},
  {Siess}, \& {Campbell}}]{doherty2014}
{Doherty} C.~L., {Gil-Pons} P., {Lau} H.~H.~B., {Lattanzio} J.~C., {Siess} L.,
  {Campbell} S.~W., 2014, \mnras, 441, 582

\bibitem[{{Doherty} {et~al.}(2015){Doherty}, {Gil-Pons}, {Siess}, {Lattanzio},
  \& {Lau}}]{doherty2015}
{Doherty} C.~L., {Gil-Pons} P., {Siess} L., {Lattanzio} J.~C., {Lau} H.~H.~B.,
  2015, \mnras, 446, 2599

\bibitem[{{Goerres} {et~al.}(1989){Goerres}, {Wiescher}, \&
  {Rolfs}}]{goerres1989}
{Goerres} J., {Wiescher} M., {Rolfs} C., 1989, \apj, 343, 365

\bibitem[{{Gratton} {et~al.}(2011){Gratton}, {Lucatello}, {Carretta},
  {Bragaglia}, {D'Orazi}, \& {Momany}}]{gratton2011hb2808}
{Gratton} R.~G., {Lucatello} S., {Carretta} E., {Bragaglia} A., {D'Orazi} V.,
  {Momany} Y.~A., 2011, \aap, 534, A123

\bibitem[{{Gratton} {et~al.}(2012){Gratton}, {Villanova}, {Lucatello},
  {Sollima}, {Geisler}, {Carretta}, {Cassisi}, \&
  {Bragaglia}}]{gratton2012sgbs1851}
{Gratton} R.~G., {Villanova} S., {Lucatello} S., {Sollima} A., {Geisler} D.,
  {Carretta} E., {Cassisi} S., {Bragaglia} A., 2012, \aap, 544, A12

\bibitem[{{Gutierrez} {et~al.}(1996){Gutierrez}, {Garcia-Berro}, {Iben},
  {Isern}, {Labay}, \& {Canal}}]{gutierrez1996}
{Gutierrez} J., {Garcia-Berro} E., {Iben} Jr. I., {Isern} J., {Labay} J.,
  {Canal} R., 1996, \apj, 459, 701

\bibitem[{{Hale} {et~al.}(2004){Hale}, {Champagne}, {Iliadis}, {Hansper},
  {Powell}, \& {Blackmon}}]{hale2004}
{Hale} S.~E., {Champagne} A.~E., {Iliadis} C., {Hansper} V.~Y., {Powell} D.~C.,
  {Blackmon} J.~C., 2004, Phys.Rev.C, 70, 045802

\bibitem[{{Han} {et~al.}(2009){Han}, {Lee}, {Joo}, {Sohn}, {Yoon}, {Kim}, \&
  {Lee}}]{han2009}
{Han} S.-I., {Lee} Y.-W., {Joo} S.-J., {Sohn} S.~T., {Yoon} S.-J., {Kim} H.-S.,
  {Lee} J.-W., 2009, \apjl, 707, L190

\bibitem[{{Iben} \& {Renzini}(1983)}]{ir1983}
{Iben} Jr. I., {Renzini} A., 1983, \araa, 21, 271

\bibitem[{{Iben} {et~al.}(1997){Iben}, {Ritossa}, \&
  {Garc{\'{\i}}a-Berro}}]{ibenritossa1997}
{Iben} Jr. I., {Ritossa} C., {Garc{\'{\i}}a-Berro} E., 1997, \apj, 489, 772

\bibitem[{{Iben} \& {Tutukov}(1984)}]{ibentutukov1984}
{Iben} Jr. I., {Tutukov} A.~V., 1984, \apjs, 54, 335

\bibitem[{{Ivans} {et~al.}(1999){Ivans}, {Sneden}, {Kraft}, {Suntzeff},
  {Smith}, {Langer}, \& {Fulbright}}]{ivans1999}
{Ivans} I.~I., {Sneden} C., {Kraft} R.~P., {Suntzeff} N.~B., {Smith} V.~V.,
  {Langer} G.~E., {Fulbright} J.~P., 1999, \aj, 118, 1273

\bibitem[{{Johnson} {et~al.}(2009){Johnson}, {Pilachowski}, {Michael Rich}, \&
  {Fulbright}}]{johnson2009}
{Johnson} C.~I., {Pilachowski} C.~A., {Michael Rich} R., {Fulbright} J.~P.,
  2009, \apj, 698, 2048

\bibitem[{{Lee} {et~al.}(2009){Lee}, {Kang}, {Lee}, \& {Lee}}]{lee2009nature}
{Lee} J.-W., {Kang} Y.-W., {Lee} J., {Lee} Y.-W., 2009, \nat, 462, 480

\bibitem[{{Lim} {et~al.}(2015{\natexlab{a}}){Lim}, {Han}, {Lee}, {Roh}, {Sohn},
  {Chun}, {Lee}, \& {Johnson}}]{lim2015b}
{Lim} D., {Han} S.-I., {Lee} Y.-W., {Roh} D.-G., {Sohn} Y.-J., {Chun} S.-H.,
  {Lee} J.-W., {Johnson} C.~I., 2015{\natexlab{a}}, \apjs, 216, 19

\bibitem[{{Lim} {et~al.}(2015{\natexlab{b}}){Lim}, {Han}, {Roh}, \&
  {Lee}}]{lim2015calcium}
{Lim} D., {Han} S.-I., {Roh} D.-G., {Lee} Y.-W., 2015{\natexlab{b}},
  Publication of Korean Astronomical Society, 30, 255

\bibitem[{{Madau} {et~al.}(1998){Madau}, {della Valle}, \&
  {Panagia}}]{madau1998}
{Madau} P., {della Valle} M., {Panagia} N., 1998, \mnras, 297, L17

\bibitem[{{Mannucci} {et~al.}(2005){Mannucci}, {Della Valle}, {Panagia},
  {Cappellaro}, {Cresci}, {Maiolino}, {Petrosian}, \& {Turatto}}]{mannucci2005}
{Mannucci} F., {Della Valle} M., {Panagia} N., {Cappellaro} E., {Cresci} G.,
  {Maiolino} R., {Petrosian} A., {Turatto} M., 2005, \aap, 433, 807

\bibitem[{{Marino} {et~al.}(2015){Marino}, {Milone}, {Karakas}, {Casagrande},
  {Yong}, {Shingles}, {Da Costa}, {Norris}, {Stetson}, {Lind}, {Asplund},
  {Collet}, {Jerjen}, {Sbordone}, {Aparicio}, \& {Cassisi}}]{marino2015}
{Marino} A.~F., {Milone} A.~P., {Karakas} A.~I., {Casagrande} L., {Yong} D.,
  {Shingles} L., {Da Costa} G., {Norris} J.~E., {Stetson} P.~B., {Lind} K.,
  {Asplund} M., {Collet} R., {Jerjen} H., {Sbordone} L., {Aparicio} A.,
  {Cassisi} S., 2015, \mnras, 450, 815

\bibitem[{{Marino} {et~al.}(2012{\natexlab{a}}){Marino}, {Milone}, {Piotto},
  {Cassisi}, {D'Antona}, {Anderson}, {Aparicio}, {Bedin}, {Renzini}, \&
  {Villanova}}]{marino2012cno}
{Marino} A.~F., {Milone} A.~P., {Piotto} G., {Cassisi} S., {D'Antona} F.,
  {Anderson} J., {Aparicio} A., {Bedin} L.~R., {Renzini} A., {Villanova} S.,
  2012{\natexlab{a}}, \apj, 746, 14

\bibitem[{{Marino} {et~al.}(2009){Marino}, {Milone}, {Piotto}, {Villanova},
  {Bedin}, {Bellini}, \& {Renzini}}]{marino2009}
{Marino} A.~F., {Milone} A.~P., {Piotto} G., {Villanova} S., {Bedin} L.~R.,
  {Bellini} A., {Renzini} A., 2009, \aap, 505, 1099

\bibitem[{{Marino} {et~al.}(2011{\natexlab{a}}){Marino}, {Milone}, {Piotto},
  {Villanova}, {Gratton}, {D'Antona}, {Anderson}, {Bedin}, {Bellini},
  {Cassisi}, {Geisler}, {Renzini}, \& {Zoccali}}]{marino2011wcen}
{Marino} A.~F., {Milone} A.~P., {Piotto} G., {Villanova} S., {Gratton} R.,
  {D'Antona} F., {Anderson} J., {Bedin} L.~R., {Bellini} A., {Cassisi} S.,
  {Geisler} D., {Renzini} A., {Zoccali} M., 2011{\natexlab{a}}, \apj, 731, 64

\bibitem[{{Marino} {et~al.}(2014){Marino}, {Milone}, {Przybilla}, {Bergemann},
  {Lind}, {Asplund}, {Cassisi}, {Catelan}, {Casagrande}, {Valcarce}, {Bedin},
  {Cort{\'e}s}, {D'Antona}, {Jerjen}, {Piotto}, {Schlesinger}, {Zoccali}, \&
  {Angeloni}}]{marino2014hb2808}
{Marino} A.~F., {Milone} A.~P., {Przybilla} N., {Bergemann} M., {Lind} K.,
  {Asplund} M., {Cassisi} S., {Catelan} M., {Casagrande} L., {Valcarce}
  A.~A.~R., {Bedin} L.~R., {Cort{\'e}s} C., {D'Antona} F., {Jerjen} H.,
  {Piotto} G., {Schlesinger} K., {Zoccali} M., {Angeloni} R., 2014, \mnras,
  437, 1609

\bibitem[{{Marino} {et~al.}(2012{\natexlab{b}}){Marino}, {Milone}, {Sneden},
  {Bergemann}, {Kraft}, {Wallerstein}, {Cassisi}, {Aparicio}, {Asplund},
  {Bedin}, {Hilker}, {Lind}, {Momany}, {Piotto}, {Roederer}, {Stetson}, \&
  {Zoccali}}]{marino2012m22}
{Marino} A.~F., {Milone} A.~P., {Sneden} C., {Bergemann} M., {Kraft} R.~P.,
  {Wallerstein} G., {Cassisi} S., {Aparicio} A., {Asplund} M., {Bedin} R.~L.,
  {Hilker} M., {Lind} K., {Momany} Y., {Piotto} G., {Roederer} I.~U., {Stetson}
  P.~B., {Zoccali} M., 2012{\natexlab{b}}, \aap, 541, A15

\bibitem[{{Marino} {et~al.}(2011{\natexlab{b}}){Marino}, {Villanova}, {Milone},
  {Piotto}, {Lind}, {Geisler}, \& {Stetson}}]{marino2011m4}
{Marino} A.~F., {Villanova} S., {Milone} A.~P., {Piotto} G., {Lind} K.,
  {Geisler} D., {Stetson} P.~B., 2011{\natexlab{b}}, \apjl, 730, L16

\bibitem[{{McLaughlin} \& {van der Marel}(2005)}]{mclvdm2005massgc}
{McLaughlin} D.~E., {van der Marel} R.~P., 2005, \apjs, 161, 304

\bibitem[{{Mennekens} {et~al.}(2010){Mennekens}, {Vanbeveren}, {De Greve}, \&
  {De Donder}}]{mennekens2010}
{Mennekens} N., {Vanbeveren} D., {De Greve} J.~P., {De Donder} E., 2010, \aap,
  515, A89

\bibitem[{{Milone} {et~al.}(2008){Milone}, {Bedin}, {Piotto}, {Anderson},
  {King}, {Sarajedini}, {Dotter}, {Chaboyer}, {Mar{\'{\i}}n-Franch},
  {Majewski}, {Aparicio}, {Hempel}, {Paust}, {Reid}, {Rosenberg}, \&
  {Siegel}}]{milone2008}
{Milone} A.~P., {Bedin} L.~R., {Piotto} G., {Anderson} J., {King} I.~R.,
  {Sarajedini} A., {Dotter} A., {Chaboyer} B., {Mar{\'{\i}}n-Franch} A.,
  {Majewski} S., {Aparicio} A., {Hempel} M., {Paust} N.~E.~Q., {Reid} I.~N.,
  {Rosenberg} A., {Siegel} M., 2008, \apj, 673, 241

\bibitem[{{Milone} {et~al.}(2012{\natexlab{a}}){Milone}, {Marino}, {Piotto},
  {Bedin}, {Anderson}, {Aparicio}, {Cassisi}, \& {Rich}}]{milone2012}
{Milone} A.~P., {Marino} A.~F., {Piotto} G., {Bedin} L.~R., {Anderson} J.,
  {Aparicio} A., {Cassisi} S., {Rich} R.~M., 2012{\natexlab{a}}, \apj, 745, 27

\bibitem[{{Milone} {et~al.}(2015){Milone}, {Marino}, {Piotto}, {Renzini},
  {Bedin}, {Anderson}, {Cassisi}, {D'Antona}, {Bellini}, {Jerjen},
  {Pietrinferni}, \& {Ventura}}]{milone2015}
{Milone} A.~P., {Marino} A.~F., {Piotto} G., {Renzini} A., {Bedin} L.~R.,
  {Anderson} J., {Cassisi} S., {D'Antona} F., {Bellini} A., {Jerjen} H.,
  {Pietrinferni} A., {Ventura} P., 2015, ArXiv e-prints

\bibitem[{{Milone} {et~al.}(2012{\natexlab{b}}){Milone}, {Piotto}, {Bedin},
  {Cassisi}, {Anderson}, {Marino}, {Pietrinferni}, \&
  {Aparicio}}]{milone2012mf2808}
{Milone} A.~P., {Piotto} G., {Bedin} L.~R., {Cassisi} S., {Anderson} J.,
  {Marino} A.~F., {Pietrinferni} A., {Aparicio} A., 2012{\natexlab{b}}, \aap,
  537, A77

\bibitem[{{Milone} {et~al.}(2012{\natexlab{c}}){Milone}, {Piotto}, {Bedin},
  {King}, {Anderson}, {Marino}, {Bellini}, {Gratton}, {Renzini}, {Stetson},
  {Cassisi}, {Aparicio}, {Bragaglia}, {Carretta}, {D'Antona}, {Di Criscienzo},
  {Lucatello}, {Monelli}, \& {Pietrinferni}}]{milone47tuc2012}
{Milone} A.~P., {Piotto} G., {Bedin} L.~R., {King} I.~R., {Anderson} J.,
  {Marino} A.~F., {Bellini} A., {Gratton} R., {Renzini} A., {Stetson} P.~B.,
  {Cassisi} S., {Aparicio} A., {Bragaglia} A., {Carretta} E., {D'Antona} F.,
  {Di Criscienzo} M., {Lucatello} S., {Monelli} M., {Pietrinferni} A.,
  2012{\natexlab{c}}, \apj, 744, 58

\bibitem[{{Miyaji} \& {Nomoto}(1987)}]{miyajinomoto1987}
{Miyaji} S., {Nomoto} K., 1987, \apj, 318, 307

\bibitem[{{Moni Bidin} {et~al.}(2007){Moni Bidin}, {Moehler}, {Piotto},
  {Momany}, \& {Recio-Blanco}}]{monibidin2007}
{Moni Bidin} C., {Moehler} S., {Piotto} G., {Momany} Y., {Recio-Blanco} A.,
  2007, \aap, 474, 505

\bibitem[{{Mucciarelli} {et~al.}(2015){Mucciarelli}, {Bellazzini}, {Merle},
  {Plez}, {Dalessandro}, \& {Ibata}}]{mucciarelli2015}
{Mucciarelli} A., {Bellazzini} M., {Merle} T., {Plez} B., {Dalessandro} E.,
  {Ibata} R., 2015, \apj, 801, 68

\bibitem[{{Nomoto}(1982)}]{nomoto1982}
{Nomoto} K., 1982, \apj, 257, 780

\bibitem[{{Nomoto} {et~al.}(2013){Nomoto}, {Kobayashi}, \&
  {Tominaga}}]{nomoto2013ARAA}
{Nomoto} K., {Kobayashi} C., {Tominaga} N., 2013, \araa, 51, 457

\bibitem[{{Nomoto} {et~al.}(1984){Nomoto}, {Thielemann}, \&
  {Wheeler}}]{nomoto1984}
{Nomoto} K., {Thielemann} F.-K., {Wheeler} J.~C., 1984, \apjl, 279, L23

\bibitem[{{Piersanti} {et~al.}(2003){Piersanti}, {Gagliardi}, {Iben}, \&
  {Tornamb{\'e}}}]{piersanti2003}
{Piersanti} L., {Gagliardi} S., {Iben} Jr. I., {Tornamb{\'e}} A., 2003, \apj,
  583, 885

\bibitem[{{Pilachowski}(1988)}]{pilac1988}
{Pilachowski} C.~A., 1988, \apjl, 326, L57

\bibitem[{{Piotto} {et~al.}(2007){Piotto}, {Bedin}, {Anderson}, {King},
  {Cassisi}, {Milone}, {Villanova}, {Pietrinferni}, \& {Renzini}}]{piotto2007}
{Piotto} G., {Bedin} L.~R., {Anderson} J., {King} I.~R., {Cassisi} S., {Milone}
  A.~P., {Villanova} S., {Pietrinferni} A., {Renzini} A., 2007, \apjl, 661, L53

\bibitem[{{Piotto} {et~al.}(2012){Piotto}, {Milone}, {Anderson}, {Bedin},
  {Bellini}, {Cassisi}, {Marino}, {Aparicio}, \& {Nascimbeni}}]{piotto2012}
{Piotto} G., {Milone} A.~P., {Anderson} J., {Bedin} L.~R., {Bellini} A.,
  {Cassisi} S., {Marino} A.~F., {Aparicio} A., {Nascimbeni} V., 2012, \apj,
  760, 39

\bibitem[{{Piotto} {et~al.}(2015){Piotto}, {Milone}, {Bedin}, {Anderson},
  {King}, {Marino}, {Nardiello}, {Aparicio}, {Barbuy}, {Bellini}, {Brown}, \&
  {Cassisi}}]{piotto2015}
{Piotto} G., {Milone} A.~P., {Bedin} L.~R., {Anderson} J., {King} I.~R.,
  {Marino} A.~F., {Nardiello} D., {Aparicio} A., {Barbuy} B., {Bellini} A.,
  {Brown} T.~M., {Cassisi} S., 2015, \aj, 149, 91

\bibitem[{{Poelarends} {et~al.}(2008){Poelarends}, {Herwig}, {Langer}, \&
  {Heger}}]{poelarendes2008}
{Poelarends} A.~J.~T., {Herwig} F., {Langer} N., {Heger} A., 2008, \apj, 675,
  614

\bibitem[{{Portegies Zwart} \& {Yungelson}(1999)}]{pgzyungelson1999}
{Portegies Zwart} S.~F., {Yungelson} L.~R., 1999, \mnras, 309, 26

\bibitem[{{Prantzos} {et~al.}(2007){Prantzos}, {Charbonnel}, \&
  {Iliadis}}]{prantzos2007}
{Prantzos} N., {Charbonnel} C., {Iliadis} C., 2007, \aap, 470, 179

\bibitem[{{Pumo} {et~al.}(2008){Pumo}, {D'Antona}, \& {Ventura}}]{pumo2008}
{Pumo} M.~L., {D'Antona} F., {Ventura} P., 2008, \apjl, 672, L25

\bibitem[{{Renzini} {et~al.}(2015){Renzini}, {D'Antona}, {Cassisi}, {King},
  {Milone}, {Ventura}, {Anderson}, {Bedin}, {Bellini}, {Brown}, {Piotto}, {van
  der Marel}, {Barbuy}, {Dalessandro}, {Hidalgo}, {Marino}, {Ortolani},
  {Salaris}, \& {Sarajedini}}]{renzini2015}
{Renzini} A., {D'Antona} F., {Cassisi} S., {King} I.~R., {Milone} A.~P.,
  {Ventura} P., {Anderson} J., {Bedin} L.~R., {Bellini} A., {Brown} T.~M.,
  {Piotto} G., {van der Marel} R.~P., {Barbuy} B., {Dalessandro} E., {Hidalgo}
  S., {Marino} A.~F., {Ortolani} S., {Salaris} M., {Sarajedini} A., 2015,
  \mnras, 454, 4197

\bibitem[{{Roediger} {et~al.}(2014){Roediger}, {Courteau}, {Graves}, \&
  {Schiavon}}]{roediger2014}
{Roediger} J.~C., {Courteau} S., {Graves} G., {Schiavon} R.~P., 2014, \apjs,
  210, 10

\bibitem[{{Rowland} {et~al.}(2002){Rowland}, {Iliadis}, {Champagne}, \&
  {Mosher}}]{rowland2002}
{Rowland} C., {Iliadis} C., {Champagne} A.~E., {Mosher} J., 2002, in APS April
  Meeting Abstracts

\bibitem[{{Siess}(2006)}]{siess2006}
{Siess} L., 2006, \aap, 448, 717

\bibitem[{{Siess}(2010)}]{siess2010}
---, 2010, \aap, 512, A10

\bibitem[{{Siess} \& {Pumo}(2006)}]{siesspumo2006}
{Siess} L., {Pumo} M.~L., 2006, \memsai, 77, 822

\bibitem[{{Smith} {et~al.}(1996){Smith}, {Shetrone}, {Bell}, {Churchill}, \&
  {Briley}}]{smith1996}
{Smith} G.~H., {Shetrone} M.~D., {Bell} R.~A., {Churchill} C.~W., {Briley}
  M.~M., 1996, \aj, 112, 1511

\bibitem[{{Straniero} {et~al.}(2014){Straniero}, {Cristallo}, \&
  {Piersanti}}]{straniero2014}
{Straniero} O., {Cristallo} S., {Piersanti} L., 2014, \apj, 785, 77

\bibitem[{{Tauris} \& {Sennels}(2000)}]{tauris-sennels2000}
{Tauris} T.~M., {Sennels} T., 2000, \aap, 355, 236

\bibitem[{{Thielemann} {et~al.}(1986){Thielemann}, {Nomoto}, \&
  {Yokoi}}]{thielemann1986}
{Thielemann} F.-K., {Nomoto} K., {Yokoi} K., 1986, \aap, 158, 17

\bibitem[{{Thygesen} {et~al.}(2014){Thygesen}, {Sbordone}, {Andrievsky},
  {Korotin}, {Yong}, {Zaggia}, {Ludwig}, {Collet}, {Asplund}, {Ventura},
  {D'Antona}, {Mel{\'e}ndez}, \& {D'Ercole}}]{thygesen2014}
{Thygesen} A.~O., {Sbordone} L., {Andrievsky} S., {Korotin} S., {Yong} D.,
  {Zaggia} S., {Ludwig} H.-G., {Collet} R., {Asplund} M., {Ventura} P.,
  {D'Antona} F., {Mel{\'e}ndez} J., {D'Ercole} A., 2014, \aap, 572, A108

\bibitem[{{Totani} {et~al.}(2008){Totani}, {Morokuma}, {Oda}, {Doi}, \&
  {Yasuda}}]{totani2008}
{Totani} T., {Morokuma} T., {Oda} T., {Doi} M., {Yasuda} N., 2008, \pasj, 60,
  1327

\bibitem[{{Trenti} {et~al.}(2015){Trenti}, {Padoan}, \& {Jimenez}}]{trenti2015}
{Trenti} M., {Padoan} P., {Jimenez} R., 2015, \apjl, 808, L35

\bibitem[{{Tutukov} \& {Yungel'Son}(1993)}]{tutukovyungelson1993}
{Tutukov} A.~V., {Yungel'Son} L.~R., 1993, Astronomy Reports, 37, 411

\bibitem[{{van Kerkwijk} \& {Kulkarni}(1999)}]{vankerkwijk-kulkarni1999}
{van Kerkwijk} M.~H., {Kulkarni} S.~R., 1999, \apjl, 516, L25

\bibitem[{{VandenBerg} {et~al.}(2012){VandenBerg}, {Bergbusch}, {Dotter},
  {Ferguson}, {Michaud}, {Richer}, \& {Proffitt}}]{vandenberg2012}
{VandenBerg} D.~A., {Bergbusch} P.~A., {Dotter} A., {Ferguson} J.~W., {Michaud}
  G., {Richer} J., {Proffitt} C.~R., 2012, \apj, 755, 15

\bibitem[{{Ventura} {et~al.}(2009){Ventura}, {Caloi}, {D'Antona}, {Ferguson},
  {Milone}, \& {Piotto}}]{ventura18512009}
{Ventura} P., {Caloi} V., {D'Antona} F., {Ferguson} J., {Milone} A., {Piotto}
  G.~P., 2009, \mnras, 399, 934

\bibitem[{{Ventura} {et~al.}(2011){Ventura}, {Carini}, \&
  {D'Antona}}]{ventura2011}
{Ventura} P., {Carini} R., {D'Antona} F., 2011, \mnras, 415, 3865

\bibitem[{{Ventura} \& {D'Antona}(2005{\natexlab{a}})}]{ventura2005a}
{Ventura} P., {D'Antona} F., 2005{\natexlab{a}}, \aap, 431, 279

\bibitem[{{Ventura} \& {D'Antona}(2005{\natexlab{b}})}]{ventura2005b}
---, 2005{\natexlab{b}}, \aap, 439, 1075

\bibitem[{{Ventura} \& {D'Antona}(2009)}]{ventura2009}
---, 2009, \aap, 499, 835

\bibitem[{{Ventura} \& {D'Antona}(2011)}]{ventura2011sagb}
---, 2011, \mnras, 410, 2760

\bibitem[{{Ventura} {et~al.}(2012){Ventura}, {D'Antona}, {Di Criscienzo},
  {Carini}, {D'Ercole}, \& {Vesperini}}]{ventura2012ngc2419}
{Ventura} P., {D'Antona} F., {Di Criscienzo} M., {Carini} R., {D'Ercole} A.,
  {Vesperini} E., 2012, \apjl, 761, L30

\bibitem[{{Ventura} {et~al.}(2000){Ventura}, {D'Antona}, \&
  {Mazzitelli}}]{ventura2000}
{Ventura} P., {D'Antona} F., {Mazzitelli} I., 2000, \aap, 363, 605

\bibitem[{{Ventura} {et~al.}(2001){Ventura}, {D'Antona}, {Mazzitelli}, \&
  {Gratton}}]{ventura2001}
{Ventura} P., {D'Antona} F., {Mazzitelli} I., {Gratton} R., 2001, \apjl, 550,
  L65

\bibitem[{{Ventura} {et~al.}(2013){Ventura}, {Di Criscienzo}, {Carini}, \&
  {D'Antona}}]{ventura2013}
{Ventura} P., {Di Criscienzo} M., {Carini} R., {D'Antona} F., 2013, \mnras,
  431, 3642

\bibitem[{{Ventura} {et~al.}(1998){Ventura}, {Zeppieri}, {Mazzitelli}, \&
  {D'Antona}}]{ventura1998}
{Ventura} P., {Zeppieri} A., {Mazzitelli} I., {D'Antona} F., 1998, \aap, 334,
  953

\bibitem[{{Webbink}(1984)}]{webbink1984}
{Webbink} R.~F., 1984, \apj, 277, 355

\bibitem[{{Willman} \& {Strader}(2012)}]{willmanstrader2012}
{Willman} B., {Strader} J., 2012, \aj, 144, 76

\bibitem[{{Yong} {et~al.}(2009){Yong}, {Grundahl}, {D'Antona}, {Karakas},
  {Lattanzio}, \& {Norris}}]{yong2009}
{Yong} D., {Grundahl} F., {D'Antona} F., {Karakas} A.~I., {Lattanzio} J.~C.,
  {Norris} J.~E., 2009, \apjl, 695, L62

\bibitem[{{Yong} {et~al.}(2015){Yong}, {Grundahl}, \& {Norris}}]{yong2015}
{Yong} D., {Grundahl} F., {Norris} J.~E., 2015, \mnras, 446, 3319

\bibitem[{{Yong} {et~al.}(2013){Yong}, {Mel{\'e}ndez}, {Grundahl}, {Roederer},
  {Norris}, {Milone}, {Marino}, {Coelho}, {McArthur}, {Lind}, {Collet}, \&
  {Asplund}}]{yong2013}
{Yong} D., {Mel{\'e}ndez} J., {Grundahl} F., {Roederer} I.~U., {Norris} J.~E.,
  {Milone} A.~P., {Marino} A.~F., {Coelho} P., {McArthur} B.~E., {Lind} K.,
  {Collet} R., {Asplund} M., 2013, \mnras, 434, 3542

\bibitem[{{Yong} {et~al.}(2014){Yong}, {Roederer}, {Grundahl}, {Da Costa},
  {Karakas}, {Norris}, {Aoki}, {Fishlock}, {Marino}, {Milone}, \&
  {Shingles}}]{yong2014}
{Yong} D., {Roederer} I.~U., {Grundahl} F., {Da Costa} G.~S., {Karakas} A.~I.,
  {Norris} J.~E., {Aoki} W., {Fishlock} C.~K., {Marino} A.~F., {Milone} A.~P.,
  {Shingles} L.~J., 2014, \mnras, 441, 3396

\bibitem[{{Zyskind} {et~al.}(1981){Zyskind}, {Rios}, \& {Rolfs}}]{zyskind1981}
{Zyskind} J., {Rios} M., {Rolfs} C., 1981, \apjl, 243, L53

\end{thebibliography}

\label{lastpage}
\end{document}